\documentclass[aps,prb,reprint,superscriptaddress]{revtex4-1}

\usepackage[version=3]{mhchem} 
\usepackage{graphicx} 

\usepackage{braket}
\usepackage{bm} 

\begin{document}


\title{Battling Retardation And Nonlocality: The hunt For The Ultimate Plasmonic Cascade Nanolens}


\author{Jamie M. Fitzgerald}
\email{jf1914@ic.ac.uk}
\affiliation{Department of Physics, Condensed Matter Theory, Imperial College London, London SW7 2AZ, United Kingdom}
\author{Vincenzo Giannini}
\homepage{www.gianninilab.com}
\affiliation{Department of Physics, Condensed Matter Theory, Imperial College London, London SW7 2AZ, United Kingdom}
\affiliation{Instituto de Estructura de la Materia (IEM-CSIC), Consejo Superior de Investigaciones Cientı́ficas, Serrano 121, 28006 Madrid, Spain}


\date{\today}

\begin{abstract}
We perform a study of the achievable field enhancement of plasmonic cascade nanolenses in the quantum and classical regimes, and explore the limits enforced at larger sizes by retardation and at smaller sizes by nonlocality and other quantum effects. We compare the near field response of a sodium nanolens within a local and nonlocal formalism and find that the increased electron-surface scattering decreases the field enhancements by over an order of magnitude. The large parameter space is explored, within the local approximation, for a number of plasmonic metals as well as the polar dielectric \ce{SiC}. Using localised surface phonon polaritons, which can be excited in polar dielectrics, is an effective strategy for overcoming retardation due to the lower energy phonon frequency. Finally, we compare the nanolens against the more usual dimer configuration, we find that the superior geometry depends crucially on the material used, with noble metal nanolenses unlikely to offer better performance to equivalent dimers. Interestingly, \ce{SiC} nanolenses can offer a larger maximum field enhancement, upto $10^4$, compared to the corresponding dimer configuration, suggesting that future endeavours in constructing nanolenses should be based on polar dielectrics.
\end{abstract}


\maketitle


\section{Introduction}
Research in plasmonics has ushered in a new era of precise control over light in subdiffraction volumes \cite{Maier2007}. One of the most famous examples of this is the self-similar nanolens, first proposed by Li \emph{et al} in 2003 \cite{Li2003}, where a finite chain of self-similar metallic nanospheres, which support localised surface plasmons (LSPs), are used to concentrate light to a ‘hot spot’ via a cascade effect. Enhancements on the order of $10^3$ are possible, which is desirable for applications in sensing \cite{Mayer2011}, energy conversion \cite{Atwater2010}, nonlinear plasmonics \cite{Kauranen2012} and surface-enhanced Raman spectroscopy (SERS) \cite{Le2008} (with potential for single molecule detection) amongst others. Despite the complex structure there has been a number of experimental realisations \cite{Bidault2007,Kneipp2008,Ding2010,Kravets2010,Hoppener2012,Coluccio2015,Lauraacoluccio2016,Heck2017,Lloyd2017} and continuing efforts.\\
\indent The original proposition for the nanolens \cite{Li2003} was formulated within the quasistatic regime, which limits the size of the constituent nanoparticles to below a few tens of nms for plasmonics in the visible. In subsequent work \cite{Li2006,Dai2008}, full electrodynamic simulations have shown that retardation effects limit the achievable field enhancement (FE) considerably. Thus, it seems sensible to stay within the electrostatic regime when designing a nanolens, but a large size difference between the nanoparticles is essential for a strong nanolensing effect, meaning the use of small nanoparticles is desirable. Unfortunately, this is experimentally difficult to realise. Further, for nanoparticles less than about $10$ nm in size, nonlocal effects start to become important \cite{Mcmahon2009,David2011,Toscano2012,Luo2013,Teperik2013,Mortensen2014} and will shift the LSP resonance and decrease the maximum FE achievable. Nonlocality in small nanoparticles is a consequence of the finite extent of the electronic wavefunctions which results in a smeared out screening charge at metallic interfaces, rather than the infinitesimally thin layer assumed by classical electromagnetism. To describe this within the framework of classical electromagnetic methods, a spatially dispersive longitudinal dielectric function is necessary. Nonlocal effects are also important for small gaps where it leads to effective increase in the separation of the particles \cite{Teperik2013,Schnitzer2016}, also reducing the FE. For very small particle sizes further quantum effects, such as the electron spillout, may become relevant \cite{Toscano2015}. Even the atomic structure may be important to include \cite{Zhang2014}. If the system size is small enough, \emph{ab initio} quantum mechanical calculations can be performed which automatically take into account nonlocal effects, this is known as quantum plasmonics \cite{Varas2016,Fitzgerald2016}. Recently it has been shown that few-atom systems can support large FEs \cite{Zhang2014,Bursi2014,Fitzgerald2017} which has led to the concept of the \emph{quantum plasmonic nanoantenna} \cite{Fitzgerald2017perspective}. Such structures could have applications in nano-localised photochemistry where the large field gradients can induce non-dipole transitions.\\
\indent Nonlocality can be modelled via the ingenious method of Luo \emph{et al} \cite{Luo2013}, where it is mimicked, in local calculations, by the inclusion of a thin (relative to the metal's skin depth) fictitious dielectric boundary layer over the metal. This model has been shown to accurately describe the blue shift of the LSP as well as the smearing of the electric field at boundaries and allows modelling of nonlocality within the computationally simpler local framework.\\
\indent For nanoparticles smaller than the mean free path there is an additional surface scattering contribution (also known as Kreibig or Landau damping) which can be simply added to the bulk damping term according to the Matthiessen's rule. The size dependent term is usually written in the form \cite{Kreibig2013}  $ \Delta \gamma(R) = A \frac{v_F}{R}$, where $A$ is a parameter on the order of unity but there exists considerable uncertainty in its value which is a consequence not only of experimental difficulties but also size dependent contributions from other sources such as phonon-plasmon coupling \cite{Fitzgerald2017,Donati2017}, structural phase transitions \cite{Kreibig1978} and adsorbate-induced damping \cite{Pinchuk2003}. There have been a number of theoretical \cite{Molina2002,Lerme2010} and experimental \cite{Quinten1996} works on this topic. Note that the surface scattering model is only a valid picture for symmetrical systems such as the sphere \cite{Kreibig2013} and spherical shell structures \cite{Moroz2008}, and is not easily generalisable to more complicated geometries.  \\
\indent In this work we perform a study of how retardation and nonlocality limits the nanolensing effect and explore the achievable FE in the classical and quantum limits. We perform an in-depth study of sodium which, due to its simplicity and similarities to the free electron gas, make it ideal for quantum calculations. We consider the effects of nonlocality (i.e. a spatially dispersive longitudinal dielectric function due to quantum effects) and surface-scattering separately and explore the contribution of each in limiting the cascade effect. We find it is the surface scattering which severely limits large FEs in self similar spherical nanolenses rather than spatial dispersion. We consider the effects of the surface scattering and nonlocality separately but it must be understood that they are two faces of the same coin, namely a classical representation of a quantum phenomena.\\
\indent We also consider more realistic plasmonic metals: gold, silver and aluminium as well as the polar dielectric silicon carbide (\ce{SiC}) which supports localised surface phonon polaritons (SPhPs) \cite{Caldwell2015} and show how a suitable choice of material and the wavelength regime of operation can lead to extreme nanolensing. Further, we compare the nanolens system against dimers of equal of volume for silver and \ce{SiC}. Somewhat surprisingly, we find no benefit of using a silver nanolens compared to the equivalent dimer geometry. In contrast for \ce{SiC} we find a substantial benefit of using the nanolens geometry, over a range of geometrical parameters, and find massive FEs approaching $10^4$. This leads us to the conclusion that the cascade effect is more suitable for the IR/THz region, where the high quality factor of polar dielectrics and the large size range at which the quasistatic approximation holds leads to nanolens operating close to the idealisation originally envisioned by Li \emph{et al}, and FEs orders of magnitude larger than what can be achieved with metal based nanolenses.\\

\section{Results And Discussions}
\begin{figure}[h]
\includegraphics[width=8cm]{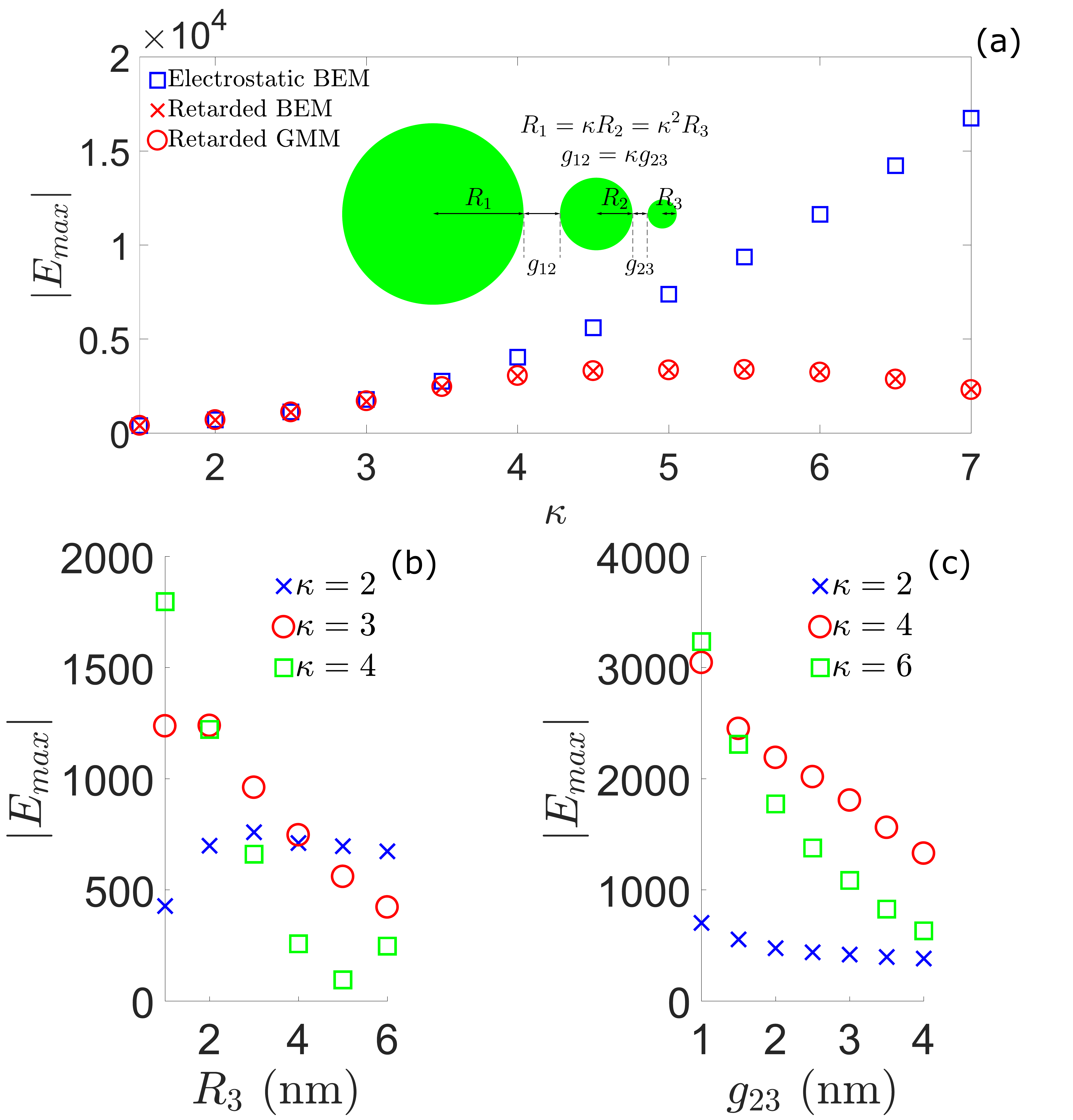}
\centering
\caption{(a) The maximum field enhancement for a sodium nanolens ($R_3=0.94$nm and $g_{23}=1$ nm) for varying $\kappa$ within the electrostatic approximation (blue) and the full retarded solution (red). As well as the boundary element method solution (crosses) also the generalised Mie theory result is show for comparison (circles). Shown inset is a schematic of the nanolens studied and the definition of the geometry parameters used in this work. (b) The maximum field enhancement for varying smallest nanoparticle radius ($R_3$) for different $\kappa$. (c) Same but for varying smallest gap size ($g_{23}$). For all these results the field is measured $0.3$ nm away from the smallest sphere surface in the gap $g_{23}$.}\label{fig:Figure_1}
\end{figure}
\subsection{Local Calculations For The Sodium Nanolens}
We begin by exploring the role of the constituent nanoparticle sizes in a nanolens within the local approximation (in this section we ignore nonlocality) using the BEM method \cite{Abajo2002,Hohenester2012,Waxenegger2015}. Despite the immense interest in the plasmonic cascade mechanism there has been few detailed studies of the huge parameter space available to modify the near field response. In the figure \ref{fig:Figure_1}a inset we show the system to be studied. We use the original geometric progression idea of Li \emph{et al} \cite{Li2003}: the smallest particle has a radius $R_3$, the medium particle has $R_2=\kappa R_3$ and the largest particle $R_1=\kappa^2 R_3$. Furthermore, we denote the smallest gap as $g_{23}$ and the larger gap will be $g_{12}=\kappa g_{23}$. The nanolens concept assumes $\kappa>>1$ so that the backcoupling of a sphere on its larger neighbour is only a small perturbation \cite{Li2003}. The largest particle acts as an antenna, coupling with the incident light via the dipole mode, and then couples with the higher order modes of the smaller spheres, which can squeeze the light into small volumes \cite{Sun2011}. We remark that we enforce this self-similar structure for simplicity but there is no guarantee that it leads to the strongest FEs possible. To explore the FE we measure the electric field $0.3$ nm away from the smallest nanoparticle in the gap between the smallest and medium nanoparticle. Unless otherwise specified we set the smallest nanoparticle to be $0.94$ nm (which corresponds to a closed shell \ce{Na92} cluster) and the smallest gap to be $1$ nm which it approximately the closest gap size one can achieve before electron tunnelling between the nanoparticles can occur (which will limit the maximum FE via short circuiting and we do not account for this, it could be included within a classical framework using the quantum corrected model \cite{Esteban2012}). In figure \ref{fig:Figure_1}a we show a sweep over $\kappa$ for both electrostatic and full retarded BEM simulations and show the maximum FE over the wavelength range $200$ nm $\rightarrow 500$ nm. Unsurprisingly for low $\kappa$ ($\lesssim 4$) the electrostatic approximation works well due to the small size of the structure, but already for $\kappa=4$ (which corresponds to the largest particle having a size $R_1=15$ nm) there is a noticeable discrepancy. For larger $\kappa$ the FE continues to grow with increasing particles size within the electrostatic approximation, in contrast the retarded calculations show a decrease in EM enhancement with increasing $\kappa$ as a result of radiative loss and shift of the resonances for larger particles. Note that these maximum electric field values may correspond to different spectral positions as the near field resonance will shift in frequency for different geometrical parameters. To confirm our results, we perform a separate calculation using the generalised Mie method (GMM) \cite{Pellegrini2007} and find excellent agreement with the BEM. The results clearly show that to achieve strong nanolensing it is desirable to simultaneously have large $\kappa$ and be within the electrostatic approximation, but this is conflicting requirements and leads to an optimum FE for a given $\kappa$. \\
\indent In figure \ref{fig:Figure_1}b we show the role of the smallest nanoparticle size ($R_3$) on the maximum FEs achievable for different $\kappa$, one can clearly see that the largest FEs are achieved for simultaneously large $\kappa$ and small $R_3$. This is a consequence of ignoring nonlocal effects and will not be the case if quantum effects are included. Interestingly, for certain values of $R_3$ one finds that larger $\kappa$ is not always advantageous. For instance, for $R_3=5$ nm the $\kappa=2$ nanolens max FE is $2.7$ times larger compared to the $\kappa=4$ case; a larger $\kappa$ is not always best! Similarly, figure \ref{fig:Figure_1}c shows the role of the gap size $g_{23}$ and reveals for larger gap sizes it may be preferable to use a smaller $\kappa$. Together these results reveal the complicated interference effects at play which are captured by the full Maxwell's equations, but not by simple electrostatic models, and illustrate the need for careful modelling of a nanolens to ensure an optimum choice of parameters is chosen.\\

\begin{figure*}[t!]
\centering
\includegraphics[width=14cm]{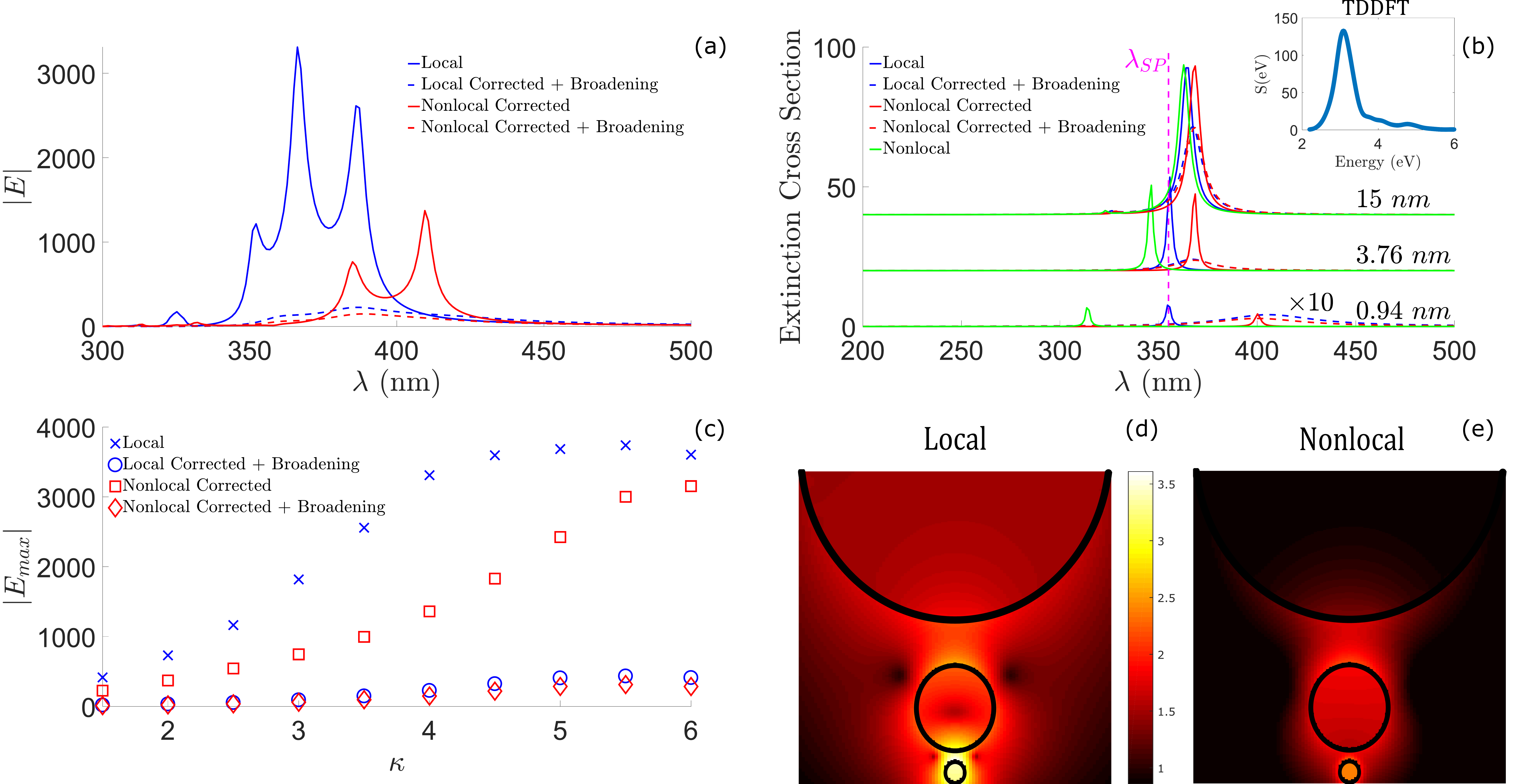}
\caption{(a) The field enhancement for a nanolens ($R_3=0.94$nm, $g_{23}=1$ nm and $\kappa=4$) within the 4 models described in the main text. (b) The extinction cross section for the individual nanospheres for different models, also shown (in green) is the result for the non-corrected nonlocal model. The local and nonlocal models with broadening included are multiplied by $10$ for clarity. The inset shows the TDDFT result for \ce{Na92} cluster, where $S(\omega)$ is the dipole strength function. (c) The maximum field enhancement for a range of $\kappa$. (d) and (e) show the field profile (logarithmic scale) within the local and nonlocal approximation at the max FE.}\label{fig:Figure_2}
\end{figure*}
\subsection{The Effect Of Nonlocality On A Sodium Nanolens}
We now include the effects of nonlocality using the local analogue model of Luo \emph{et al} \cite{Luo2013} within the BEM. Our findings in the last section revealed that it is desirable to choose a small $R_3$ and large $\kappa$, it is interesting to see how this breaks down with quantum effects included. For any practical system, a full numerical quantum calculation is clearly infeasible due to the large size so we make use of a number of approximations to model quantum effects. As the smallest sphere is well within the quantum limit ($R_3<1$ nm), here we use the spherical jellium model within the local density approximation-TDDFT formalism, which provides a good description for closed shell clusters, to describe the optical response of the \emph{individual} smallest nanosphere. This provides results accurate to within a few tenths of an electron volt in comparison to experimental results \cite{Xia2009}. We find the TDDFT result does not agree well with the nonlocal model at these sizes for individual nanoparticle optical response, the nonlocal result can be 'corrected' following the prescription of Teperik \emph{et al} \cite{Teperik2013}. Performing the TDDFT calculation allows an estimation of the electron spill out not possible relying on quasi-quantum nonlocal models and allows modelling of nanolenses down to the quantum limit.\\
\indent In figure \ref{fig:Figure_2}a we show the FE for a sodium nanolens of $R_3 = 0.94$ nm, $g_{23}=1$ nm and $\kappa = 4$ for 4 different models. The 1st model is purely local as used in the last section. The 2nd model is local but with surface scattering contribution included in the damping as well as an electron spillout correction to the plasma frequency provided by the TDDFT calculation. Note that for the smallest sphere $\ce{Na92}$ we use an experimental value of $0.42$ eV \cite{Xia2009} for the damping rather than the Kreibig formula which will breakdown at the smallest scale. Sodium is well modelled by a free electron gas and as $\omega_P>>\gamma$ so that we may take the plasmon line width to be equal to $\gamma$. The 3rd model is a nonlocal calculation with \emph{both} an electron spillout correction and a correction to fix the incorrect value of the Feibelman parameter for alkali clusters \cite{Teperik2013, Monreal2013}. The non-corrected hydrodynamic model predicts the screening charge inside the metal surface leading to a negative Feibelman parameter, this is contradiction to full quantum calculations and experimental results. The plasma frequency can be corrected as 
\begin{equation}
\omega_{SP} = \omega_{SP}^{nonloc}(1 - \frac{\Delta}{R} - \frac{\delta}{R})
\end{equation}
where $R$ is the nanosphere radius, $\Delta$ is the position of the induced charge relative to the edge and $\delta$ is the spillout length \cite{Teperik2013}. The local results can be similarly corrected $\omega_{SP} = \omega_{SP}^{loc}(1  - \frac{\delta}{R})$, which is done in model $2$. Both quantities are on the order of an $\AA$ so we will take $\delta \approx \Delta$ and fit according to the TDDFT simulation for a \ce{Na92} sodium cluster, we find that a value of $\delta = 0.12$ nm gives a good fit to the quantum simulation, this agrees fairly well with the experimental value of $0.145$ nm \cite{Reiners1995}. The 4th model is a nonlocal calculation with the correction and the surface scattering contribution to the damping. Note that the $\delta$ parameter is calculated from the single \ce{Na92} cluster and that value is used for the correction for \emph{all} the spheres in the nanolens.\\
\indent By considering these 4 models we can explore the contribution from the various small size effects on the FE. In the local model, we find a number of peaks due to a complicated plasmon hybridisation between the 3 particles. Interestingly, compared to the individual nanosphere response, the large FE response is rather broadband; over a range of about $50$ nm a large FE of over $1000$ is possible. It is also worth remembering that we are only recording the field at one point so there may be large FEs at other points not captured by these results. We find that the nonlocal model with no broadening leads to a redshift (due to electron spillout) and a reduction in the maximum FE to about $40 \%$ of the local result, there is also a smoothing out of the number of peaks visible. For both model $2$ and $4$ there is a serious reduction in the FE by over a factor of $10$ due to Landau damping, such a reduction means that the nanolens gives no benefit over ordinary individual nanospheres and dimers which offer FEs on the order of $Q$ and $Q^2$ respectively \cite{Sun2011}, where $Q$ is the quality factor and is on the order of $10$ for typical plasmonic metals near the LSP resonance. This leads us to conclude it is the increased damping via surface scattering that severely limits the cascade FE rather than the nonlocal shift of the resonances.\\
\indent To further understand these results we can analyse the optical response of each individual nanosphere. In figure \ref{fig:Figure_2}b we show the extinction cross section \cite{Bohren2008}, for each model, of the individual nanoparticles in the $\kappa=4$ system. Also shown (in green) is the non-corrected nonlocal model for comparison. We can see for the largest particle ($R=15$ nm) that, somewhat surprisingly, retardation is already important and shifts the resonance to lower energies, at smaller wavelengths we can see a weak higher order mode beginning to form. At all 3 sizes, we can see that the non-corrected nonlocal incorrectly predicts a blueshift. The corrected models in contrast show a redshift which is a consequence of the electron spillout, naturally this shift is smaller for larger nanoparticles. For some other metals, such as silver, the nonlocal shift is towards higher energies due to the dynamical screening of d-electrons \cite{Monreal2013} and the non-corrected nonlocal model will be more accurate, although this should be regarded as a lucky coincidence. It is interesting that for the $15$ nm sodium nanoparticle that both retardation and quantum effects have a visible effect, this hints there is a size regime where fully retarded and quantum calculations are necessary using models such as the recently developed quantum hydrodynamic model \cite{Yan2015}. In the inset of \ref{fig:Figure_2}b we show the jellium TDDFT result for \ce{Na92} where $S$ is the dipole strength function, we use a artificial $0.1$ eV broadening. We find close to $3$ eV a prominent LSP is present, at lower broadening it is possible to see that in fact the peak is fragmented via interactions with single particle excitations. Also present is a Bennett surface plasmon and the volume plasmon at higher energies \cite{Varas2016}, these modes are not included in the local and nonlocal models used. The SPP is red shifted compared to the classical result due to the soft confining potential.\\
\indent In figure \ref{fig:Figure_2}c we show the max FE for the 4 different models over a range of $\kappa$. We see that the trends from figure \ref{fig:Figure_2}a continue for different values of $\kappa$. Interestingly at larger $\kappa$ the surface scattering role in decreasing the FE is increased, this is a consequence of the smaller particle not contributing to the cascade effect whilst the larger $2$ spheres begin to behave purely classical; in effect, the system behaves as an asymmetric dimer and we can expect field enhancements on the order of $Q^2$ rather than $Q^3$ if surface scattering is not included \cite{Li2003}. We also show a logarithmic plot of the FE for the max FE wavelength at $\kappa=4$ within the local (model 1) and nonlocal corrected (model 4) in figure \ref{fig:Figure_2}d and \ref{fig:Figure_2}e respectively. The local result shows that the largest field enhancement is near the smallest sphere in the gap $g_{23}$, this agrees with what was found in the original work on the nanolens \cite{Li2003}. In contrast, the FE spatial profile for the nonlocal models shows only a small FE in the gap with the largest fields found solely within the smallest nanoparticle, this illustrates a breakdown of the cascade effect. \\
\indent It seems from these results that going smaller is not an effective strategy for building plasmonic cascade devices. Recently, similarly drastic reductions (up to $7$ times) from nonlocality have been shown for the fluorescence enhancement of a dipole near a gold nanoparticle \cite{Jurga2017}. We emphasise that the inclusion of electron tunnneling, a neglible effect at these gap sizes, would only further limit the field ehancement. It is worth stressing that nonlocality is highly dependent on geometry, for instance thin metallic nanoshells could offer superior performance for ultrasmall nanolenses. Experimental results have shown that $20$ nm thick gold nanoshells shows no additional broadening \cite{Nehl2004}. This has been backed up by theoretical calculations within the random phase approximation which have shown Landau damping decreases with decreasing nanoshell thickness \cite{Kirakosyan2016}.\\

\begin{figure}[h!]
\includegraphics[width=8cm]{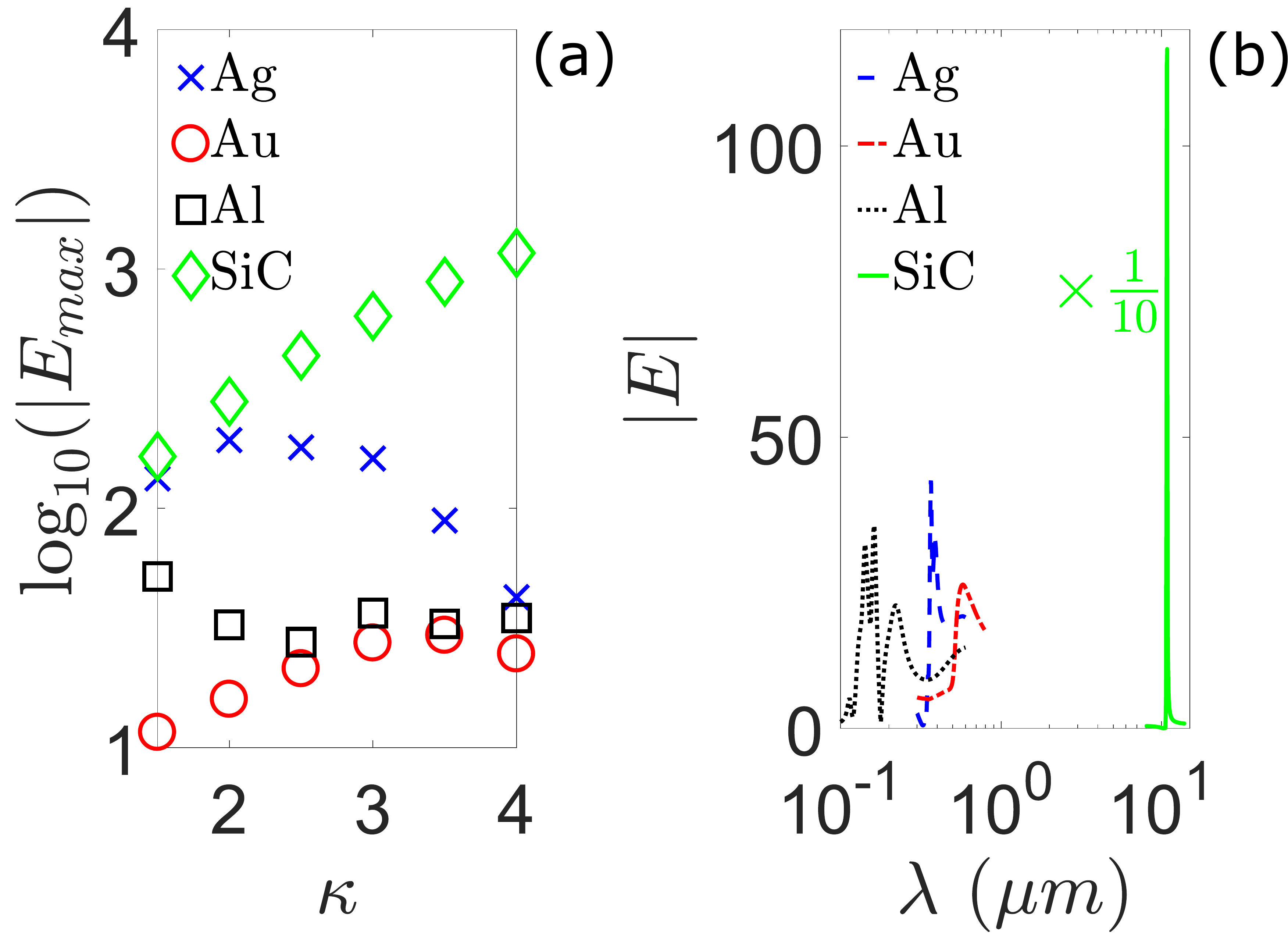}
\centering
\caption{(a) Maximum field enhancement, in log scale, for silver, gold, aluminium and silicon carbide for different $\kappa$. the geometry is $R_3 = 5$ nm, $g_{23} = 3$ nm. (b) The field enhancement against wavelength for the $\kappa=4$ geometry, with the \ce{SiC} result multiplied by $1/10$ to allow visualisation.}\label{fig:Figure_3}
\end{figure}
\subsection{Local Calculations for Gold, Silver, Aluminium and Silicon Carbide}
In the preceding sections, we have seen the detrimental effects of retardation and nonlocality on the plasmonic cascade effect. To explore different strategies for achieving large FEs we now consider the effect of the material. We study the effects of changing $\kappa$ for more typical plasmonic metals: silver, gold and aluminium. We also look at the polar dielectric \ce{SiC} which, within the Reststrahlen band that is located in the mid-IR between the longitudinal and transverse optical phonons, behaves as an optical metal and supports LSPhPs \cite{Caldwell2015}. The low loss and high Q (which can be over an order of magnitude larger than for noble metals, this is because the non-radiative loss is determined by phonon-phonon scattering rather than electron-electron scattering in metals) makes polar dielectrics suitable for use in a IR/THz nanolens. Here we study the maximum FE over a range of $\kappa$, for simplicity we use a local model as we now set $R_3=5$ nm and $g_{23}=3$ nm which are easier for experimental realisation and are similar to the geometry used in the original paper by Li \emph{et al} \cite{Li2003}. In figure \ref{fig:Figure_3} we show the maximum FE $0.3$ nm away from the smallest nanosphere. We find that gold, the most commonly used plasmonic metal, has the lowest maximum FE: always below $100$. This is a consequence of interband transitions close to the plasmon resonance which limits the quality factor. A more optimum nanolens could be constructed for gold by structuring (for instance into an ellipsoid or rod shape) so that that the LSP is shifted to larger wavelengths. Aluminium exhibits higher FEs and is useful for achieving large FEs in the UV. Silver is good choice for achieving large FEs due to its large Q for a metal (approximately an order of magnitude larger than gold at the plasmon frequency \cite{Caldwell2015}). For low $\kappa$ ($\lesssim 2.5$) it has the largest FEs of the 4 materials studied. For larger $\kappa$, \ce{SiC} shows the largest FE due to its high Q combined with a lack of retardation loss. In contrast to the monotonically increasing FE with $\kappa$ for \ce{SiC}, the metals show a more complicated variation with $\kappa$ (similar to what is seen for sodium in figure \ref{fig:Figure_1}a), which is a hallmark of retardation effects. This is confirmed by comparison with an electrostatic calculation for \ce{SiC}, which is very similar to the full retarded results. This is a consequence of the low frequency of the \ce{SiC} LSPhP as compared to the LSP of the metals. The electrostatic approximation holds for a nanosphere, at the resonance, if $|\sqrt{\epsilon}|2 \pi R/\lambda_{SPhP} <<1$ \cite{Bohren2008} which for \ce{SiC}is found to be around a micron, hence for $\kappa=4$ where the largest sphere radius is $80 $ nm this is very well satisfied still. In comparison \ce{Al}, which has a high frequency SPP, very quickly deviates from the electrostatic approximation for very small particle sizes. This suggests that the nanolens concept is more viable for applications in the IR/THz where the electrostatic approximation holds at larger sizes. Note that the maximum FE is highly sensitive to the material data used as shown in reference \cite{Pellegrini2016}, where the maximum FE was shown to change by a factor of $5$ depending on the experimental data used.\\
\indent  The use of SPhPs for the nanolens cascade effect offers an alternative to using LSPs and should lead to the larger FEs, although one must work at longer wavelengths in the IR/THz. Fortunately, this is a window of the electromagnetic spectrum that has generated a huge amount of interest, it coincides with vibrational and rotational transitions of molecules \cite{Caldwell2015}. Similar results may also be obtained in this spectral region for doped semiconductors \cite{Luther2011} and graphene \cite{Koppens2011}. The FE values we have obtained for the \ce{SiC} nanolens are upper bounds as we have not included any nonlocal or surface scattering corrections, we expect for the systems studied here that such effects will be small and should compare favourably to metals. The flat dispersion of optical phonon in the long wavelength limit leads to a low group velocity on the order of $ \sim 10^4$ m/s for polar crystals, this gives, despite the relatively long scattering time of around $ \sim 10^2$ ps, a short mean free path on the order of $10$ nm. Furthermore, quantum corrections can be expected to be negligible due to an absence of free carriers.

\begin{figure}[h!]
\includegraphics[width=8cm]{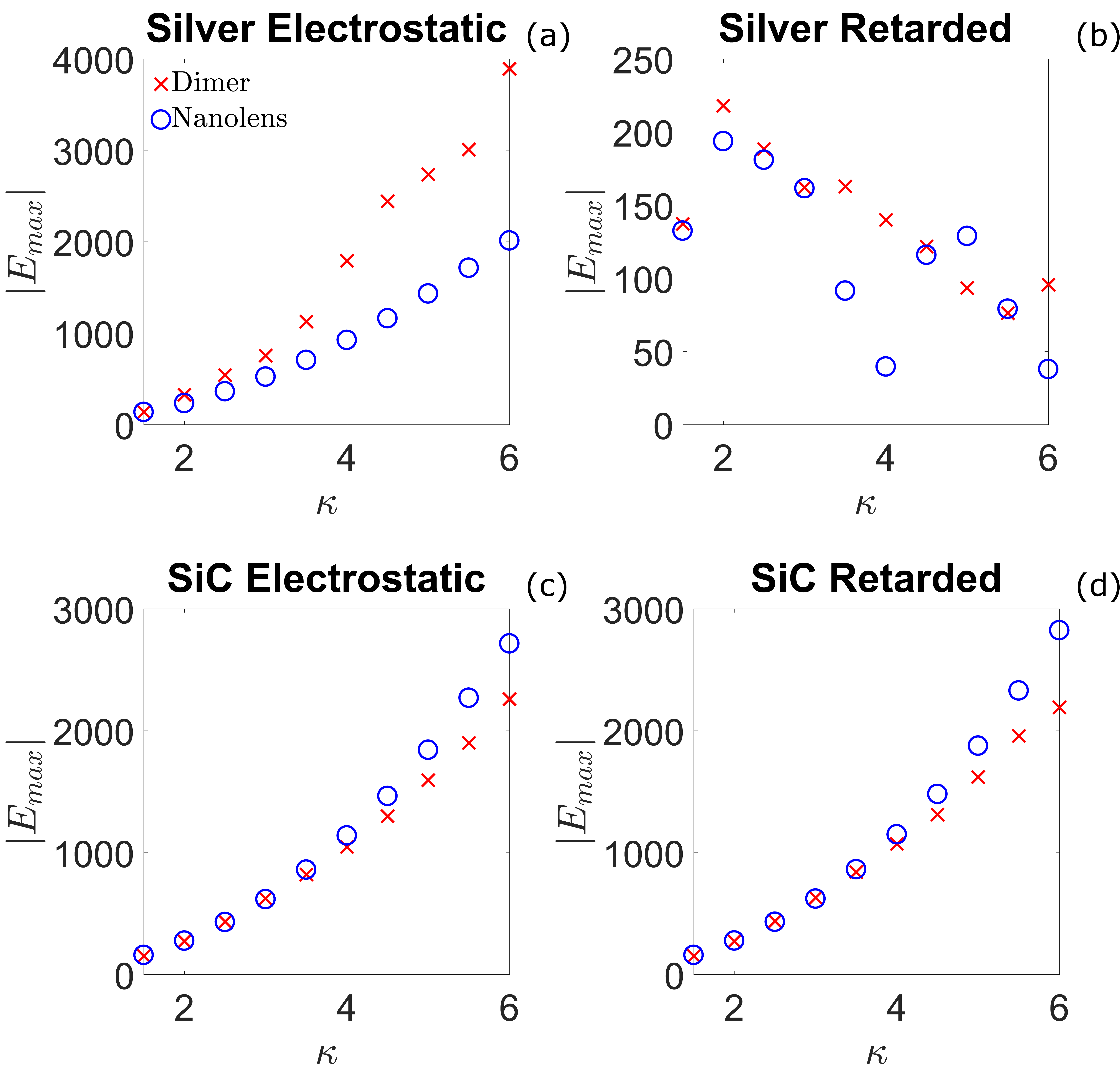}
\centering
\caption{The maximum field enhancement of silver nanolens and dimers for different $\kappa$ calculated within the electrostatic approximation (a) and full retarded solution (b). The geometry parameters are $R_3=5$ nm and $g_{23}=3$ nm. (c),(d) Same but for silicon carbide.}\label{fig:Figure_4}
\end{figure}
\subsection{Nanolens Vs Dimer}
\begin{figure}[t]
\includegraphics[width=8cm]{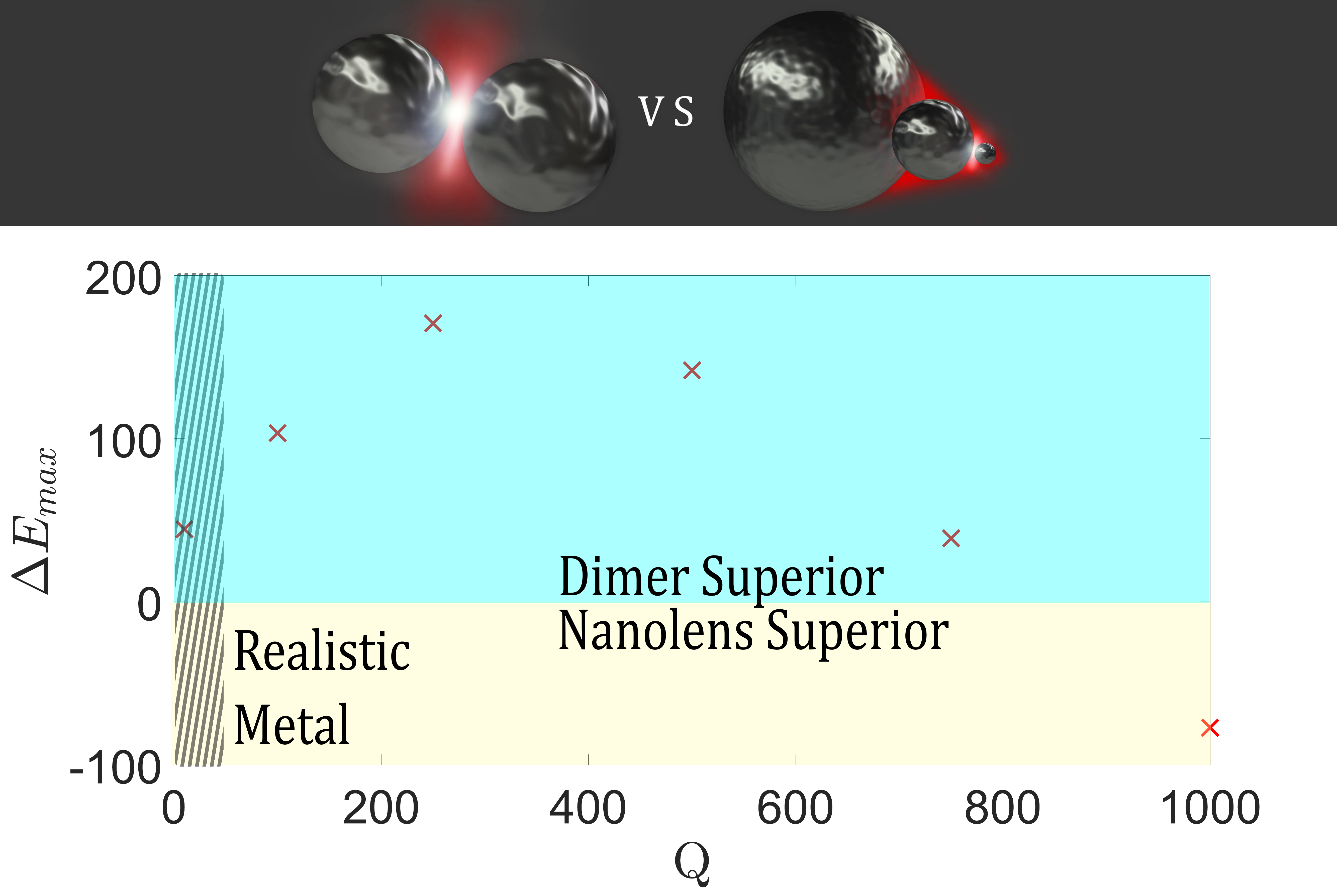}
\centering
\caption{The maximum field enhancement of a dimer minus that of a nanolens for a fictitious Drude metal with variable $Q$. The geometry parameters are $R_3=5$ nm, $g_{23} = 3$ nm and $\kappa=4$. Indicated in blue are the regions where the dimer is superior and cream are the region where the nanolens is superior. The hashed region indicated the region of realistic Q values for metals.}\label{fig:Figure_6}
\end{figure}

\begin{figure}[t]
\includegraphics[width=8cm]{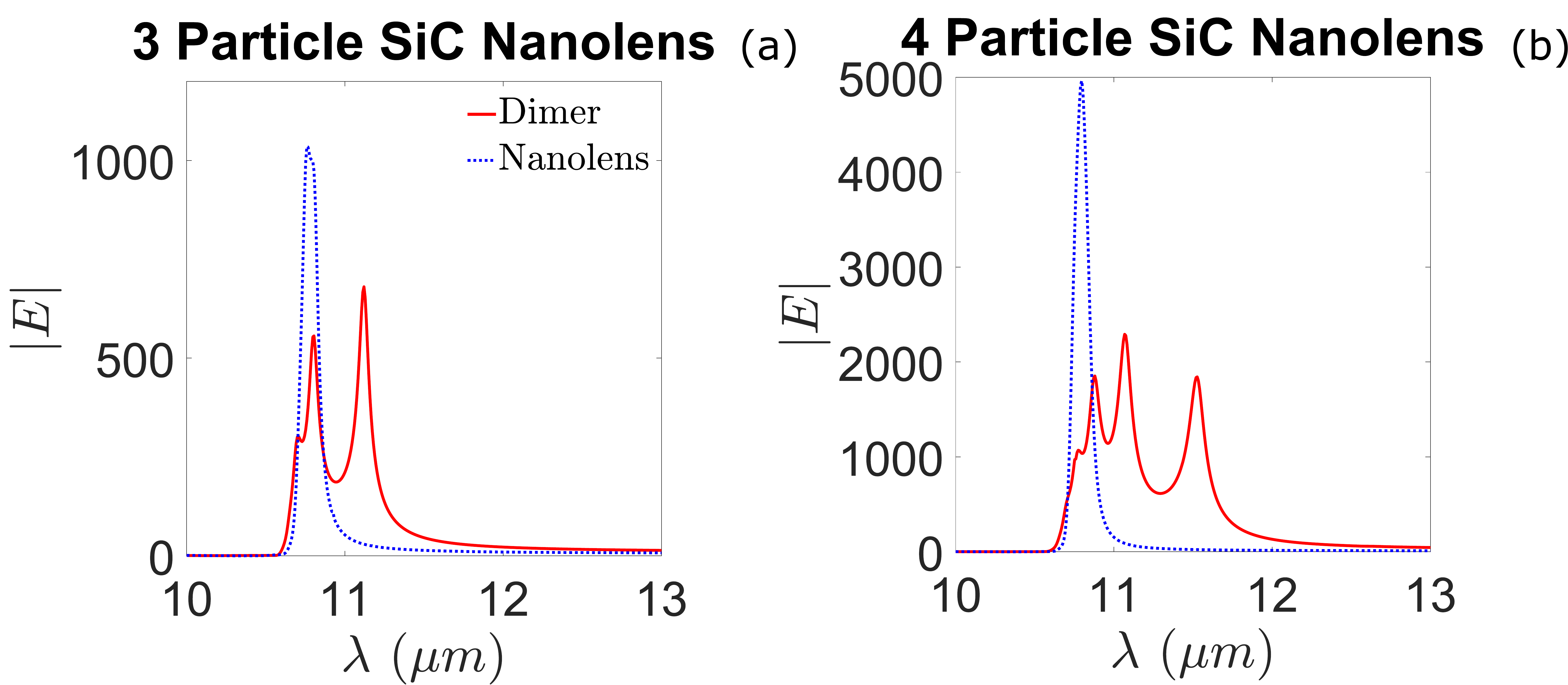}
\centering
\caption{The field enhancement for: (a) $3$ and (b) $4$ particle \ce{SiC} nanolens, with the smallest nanosphere radius and gap equal to $10$nm, and $\kappa=4$. Both are compared with the equivalent volume dimer.}\label{fig:Figure_7}
\end{figure}

Finally, we comment on the suitability of the cascade effect to achieve large FEs in comparison to the more usual dimer setup. Recently it was suggested by Pellegrini \emph{et al} \cite{Pellegrini2016} that there are no significant improvements achieved with self-similar nanolenses as compared to plasmonic dimers of equal or less total volume. This was a surprising result but was only tested for a single nanolens geometry. Here we explore this in detail for a range of $\kappa$ for both silver and \ce{SiC}. For a fair comparison we enforce the dimer volume to be equal to be the same as the equivalent nanolens for a given $\kappa$ so that $R_{dimer} = ((R_1^3+R_2^3+R_3^3)/2)^{1/3}$ and the gap is the same as the smallest gap ($g_{23}$) of the nanolens. We measure the field in the middle of the gap for the dimer.\\
\indent In figure \ref{fig:Figure_4}a and b is shown the maximum FE over a range of $\kappa$ for a silver nanolens and the equivalent volume symmetric dimer, for the electrostatic approximation and full retarded solution respectively. We find for silver that the findings of Pellegrini \emph{et al} hold for all geometries considered in the electrostatic approximation and all, expect at $\kappa=5$, for the full retarded calculation. Our work indicates that the nanolens geometry holds no significant advantage over the dimer system in the case of silver for exciting strong FEs. Whilst more work is needed over a larger parameter range and different metals, this does suggest that the huge amount of work in search for large field enhancements with nanolens built of noble metals may be wasted effort when much simpler dimers are preferable. What is most surprising, also noticed by Pellegrini \emph{et al} \cite{Pellegrini2016}, is that the dimer remains superior within the electrostatic approximation as well (figure \ref{fig:Figure_4}a). Thus, it is not retardation effects that limit the nanolensing effect, rather it is a limitation of the material.\\
\begin{figure}[t]
\includegraphics[width=8cm]{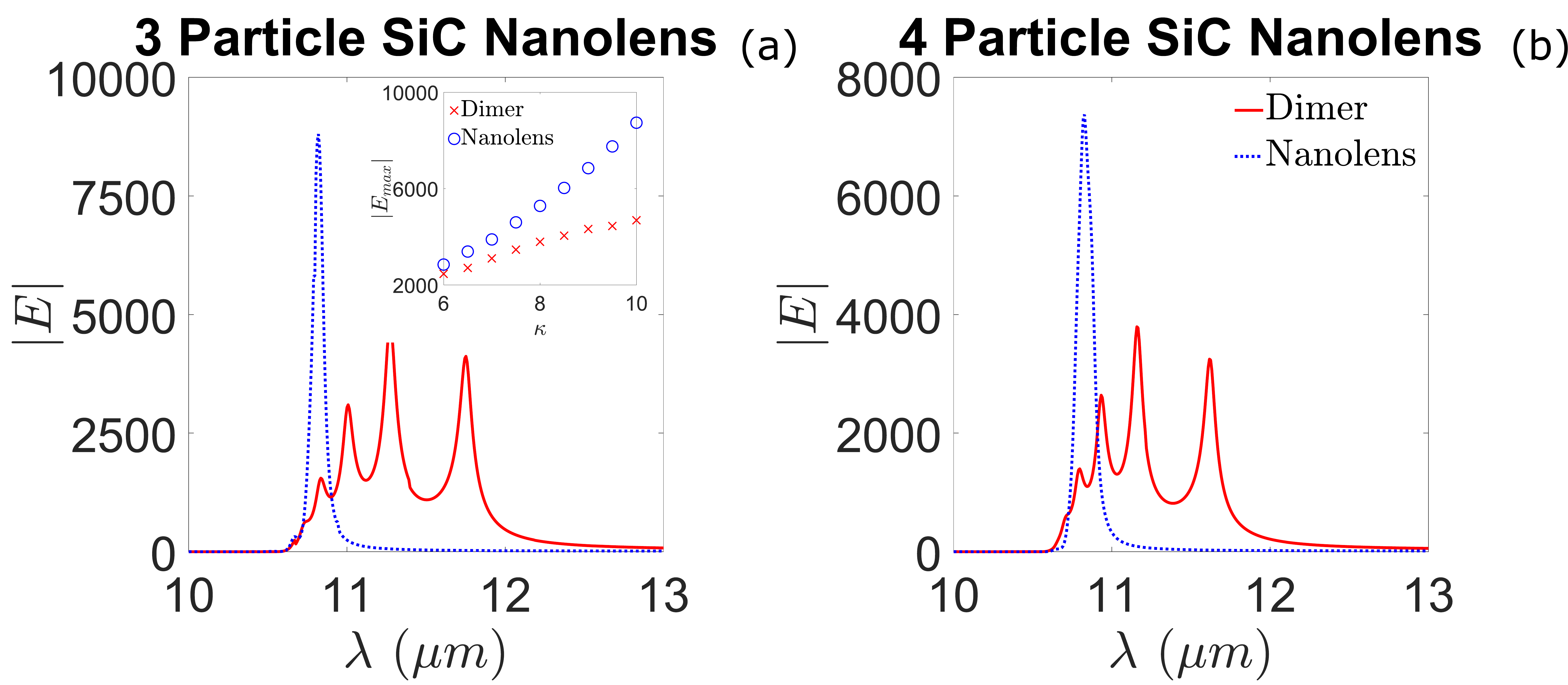}
\centering
\caption{The field enhancement for:  (a) a 3 particle \ce{SiC} nanolens of geometry $R_3=5$nm, $g_{23}=3$nm and $\kappa=10$, (b) a 4 particle nanolens of geometry $R_4=5$nm, $g_{34}=3$nm and $\kappa=4$. Both are compared with the equivalent volume dimer.}\label{fig:Figure_5}
\end{figure}
\indent In contrast, for \ce{SiC} (see figure \ref{fig:Figure_4}c and d) we find that the nanolens become increasingly superior to the dimer for larger $\kappa$, at $\kappa=6$ the nanolens geometry leads to a $29 \%$ larger FE. The suitability of SiC for nanolensing is a consequence of being well within the electrostatic limit at these sizes, combined with low material loss (large Q). The former being confirmed by the electrostatic results (figure \ref{fig:Figure_4}c) being approximately equal to the retarded results (figure \ref{fig:Figure_4}d). Note that we expect the nanolens to be further superior to the dimer for larger $\kappa$ than shown in figure \ref{fig:Figure_4}, where we restrict ourselves to $\kappa \leq 6$; the BEM calculations become increasingly difficult to converge as the size ratio between the constituent spheres increases. To confirm this intuition we have performed an additional calculation (using the multi-sphere T-matrix code by Mackowski and Mishchenko \cite{Mackowski2011}, which is able to achieve convergence for these challenging geometries) for a $\kappa=10$ nanolens (see figure \ref{fig:Figure_5}a) and find the nanolens geometry has a $82 \%$ improvement of the maximum FE compared to the equivalent volume dimer. By using \ce{SiC} structures we can approach massive FEs of $10^4$, which corresponds to an intensity enhancement of $10^8$. This demonstrates that polar dielectrics are a suitable material for constructing extreme-cascade nanophotonic devices. The inset of figure \ref{fig:Figure_5}a shows the maximum FE for $\kappa$ from $6$ to $10$ for both the dimer and nanolens and shows that the nanolens become increasingly superior for larger $\kappa$, continuing the trend from the BEM calculation shown in \ref{fig:Figure_4}d. The drop off in FE increment with increasing $\kappa$ for the dimer is presumably due to growing retardation loss, for large enough $\kappa$ a similar drop off will be seen for the nanolens. Interestingly, the spectral information in figure \ref{fig:Figure_5}a shows that the strongest FE occurs for a narrow single peak and we have found this to be an hallmark of strong nanolensing, this is in contrast to the dimer which has multiple peaks due to mode hybridisation. We have confirmed that the response of the system is down to the material resonance rather than a pure geometric resonance by calculating the FE for a silver nanolens in the same wavelength regime as for the \ce{SiC} nanolens ($8 \mu m \rightarrow 14 \mu m$) where the silver acts, to a good approximation, as a perfect electrical conductor, and we find only a small FE on the order of $10$. \\
\indent To understand the results shown in figure \ref{fig:Figure_4} further, we model a fictitious Drude model (based on silver) and vary the quality factor, for $\omega>>\gamma$, this is as simple as changing $\gamma$. We then plot the maximum FE of the dimer minus the nanolens for $\kappa=4$ (see figure \ref{fig:Figure_6}). Negative number correspond to the nanolens outperforming the dimer. The results show that the nanolens geometry, for these particular parameters, is desirable when the quality factor is very large ($Q \gtrsim 800$) which is far beyond what is achievable in plasmonics in the visible (the Q of silver at the LSP resonance is $\sim 30$) and is only just within the reach of the best of polar dielectrics at much lower frequencies (\ce{SiC} has $Q \sim 900$ at the LSPhP resonance) \cite{Caldwell2015}. Alternative materials suitable for constructing nanolens could be high index dielectrics which exhibit very large quality factors or hybrid dielectric-metal systems where the loss can be modified over orders of magnitude ($\sim 10^3$) from metal to dielectric-like \cite{Yang2017}.\\
\indent  It is worth mentioning that in the original work of  Li \emph{et al} it was shown that a symmetric nanolens increases the FE by a factor of $2$. So far in this work, we have stuck to $3$ self-similar nanospheres but higher FEs can be achieved with a larger number of elements, although at the cost of added complexity to build. Such devices have been experimentally demonstrated \cite{Coluccio2015} and lead to improved SERS intensity \cite{Lauraacoluccio2016} compared to the 3 particles nanolens. In figure \ref{fig:Figure_5}b we show the results for a nanolens built of $4$ spheres. The smallest sphere radius ($R_4$) and gap ($g_{34}$) are $5$nm and $3$nm respectively and $\kappa=4$. We find that the nanolens can achieve a maximum FE of almost double the equivalent dimer, again demonstrating the effectiveness of the cascade effect in SiC devices. Higher numbers of nanospheres could be considered and could well lead to even larger improvements.\\
\indent As a final remark in this section, we note that the geometries shown, whilst demonstrating extreme FE, may be difficult to produce for experimental demonstration due to the large $\kappa$ and small gaps. These results should be taken as an indication of the ultimate achievable FEs in polar dielectrics (although we emphasise that further geometrical optimisation is certainly possible). To demonstrate a more attainable device, we consider a nanolens with both the smallest nanosphere radius and gap to be $10 nm$ and limit ourselves to $\kappa=4$, which is inline with what is experimentally achievable. The results are shown in figure \ref{fig:Figure_7}, we find for these geometries that the maximum FE is $52\%$ and $116 \%$ larger for the $3$ and $4$ particle nanolens geometry, as compared to the equivalent dimer, respectively. The value of the FE is, of course, lower than the results shown in figure \ref{fig:Figure_5} due to the larger gap. A wider spacing leads to a lower number of higher order plasmon modes being excited and a consequent drop in the field concentration near the smaller nanosphere. The results are clear evidence that \emph{experimentally realistic} \ce{SiC} devices can utilise the cascade effect to achieve large FEs, beyond what is achievable with metal based devices.

\section{Conclusions}
We have shown that nonlocal effects hugely limit the possible FE in plasmonics nanolens and have shown that building smaller metallic nanolens to beat retardation is not a viable strategy because of nonlocality. Our \emph{top-down} approach using quasi-quantum models contrasts with existing studies, based on density functional theory which have found suprisingly large FEs \cite{Zhang2014,Bursi2014,Fitzgerald2017} and emphasises the need for a greater understanding of the loss channels relevant to plasmonics at these length scales. A more promising route to achieve large FEs is to use suitable materials or geometries with resonances in the IR/THz where higher quality factors and a much larger span of $\kappa$ within the electrostatic regime, are possible. In particular we have illustrated the potential of \ce{SiC} nanolenses to achieve extreme FEs on the order of $10^4$, we can expect even higher FEs for optimised structures. These nanolenses could have applications in molecular sensing in the mid-IR range and could also be suitable for achieving coupling with graphene plasmons and molecular excitations to create tunable hybrid modes \cite{Li2017}. The use of high-index dielectrics for the cascade effect is also promising due to their much superior quality factor compared to metals.

\section{Methods}
For sodium we use a Drude dielectric function $\epsilon(\omega) = 1 - \frac{\omega_P^2}{\omega^2 + i \omega \gamma}$ with parameters $\omega_P = 6.05 $eV, $\gamma = 0.02684$ eV. The plasma frequency is determined by the Wigner-Seitz radius $2.08 \AA$ which is also used to determine the Jellium density. For gold and silver we use experimental data from Johnson and Christy \cite{Johnson1972}. For aluminium, Drude parameters of $\omega_P= 15.3$ eV and $\gamma = 0.598$ eV \cite{Blaber2009}. For \ce{SiC} we use a Lorentz oscillator model $\epsilon(\omega) = \epsilon_{\infty} (1+\frac{\omega_L^2-\omega_T^2}{\omega_T^2-\omega^2-i \omega \gamma})$ with the longitudinal and transverse optical phonon frequencies $\omega_T = 0.0988$ eV and $\omega_L = 0.120$ eV respectively, $\epsilon_{\infty} = 6.56$ and $\gamma = 0.00059$ eV \cite{Francescato2014}. Throughout, where possible, we checked local calculations by using both the BEM and GMM method and found excellent agreement.\\
\indent When using the local analogue model, we set the artificial dielectric thickness to $R/200$ throughout, we have checked the accuracy of the method by comparison to coated Mie theory for a single sphere \cite{Bohren2008}. For the TDDFT calculations we used the real space code OCTOPUS \cite{Andrade2015}. To test the validity of our simulation we checked the sum rule $\int s(\omega) d\omega = N$, where $N$ is the number of valence electrons, and found it was satisfied to $99.8\%$.

\begin{acknowledgments}
The work of J. M. Fitzgerald was supported
under a studentship from the Imperial College London funded by the EPSRC Grant 1580548. We would like to express gratitude to the authors of the open source software used in this work for making their codes freely available online.
\end{acknowledgments}

\bibliography{Bib}

\begin{thebibliography}{64}%
\makeatletter
\providecommand \@ifxundefined [1]{%
 \@ifx{#1\undefined}
}%
\providecommand \@ifnum [1]{%
 \ifnum #1\expandafter \@firstoftwo
 \else \expandafter \@secondoftwo
 \fi
}%
\providecommand \@ifx [1]{%
 \ifx #1\expandafter \@firstoftwo
 \else \expandafter \@secondoftwo
 \fi
}%
\providecommand \natexlab [1]{#1}%
\providecommand \enquote  [1]{``#1''}%
\providecommand \bibnamefont  [1]{#1}%
\providecommand \bibfnamefont [1]{#1}%
\providecommand \citenamefont [1]{#1}%
\providecommand \href@noop [0]{\@secondoftwo}%
\providecommand \href [0]{\begingroup \@sanitize@url \@href}%
\providecommand \@href[1]{\@@startlink{#1}\@@href}%
\providecommand \@@href[1]{\endgroup#1\@@endlink}%
\providecommand \@sanitize@url [0]{\catcode `\\12\catcode `\$12\catcode
  `\&12\catcode `\#12\catcode `\^12\catcode `\_12\catcode `\%12\relax}%
\providecommand \@@startlink[1]{}%
\providecommand \@@endlink[0]{}%
\providecommand \url  [0]{\begingroup\@sanitize@url \@url }%
\providecommand \@url [1]{\endgroup\@href {#1}{\urlprefix }}%
\providecommand \urlprefix  [0]{URL }%
\providecommand \Eprint [0]{\href }%
\providecommand \doibase [0]{http://dx.doi.org/}%
\providecommand \selectlanguage [0]{\@gobble}%
\providecommand \bibinfo  [0]{\@secondoftwo}%
\providecommand \bibfield  [0]{\@secondoftwo}%
\providecommand \translation [1]{[#1]}%
\providecommand \BibitemOpen [0]{}%
\providecommand \bibitemStop [0]{}%
\providecommand \bibitemNoStop [0]{.\EOS\space}%
\providecommand \EOS [0]{\spacefactor3000\relax}%
\providecommand \BibitemShut  [1]{\csname bibitem#1\endcsname}%
\let\auto@bib@innerbib\@empty
\bibitem [{\citenamefont {Maier}(2007)}]{Maier2007}%
  \BibitemOpen
  \bibfield  {author} {\bibinfo {author} {\bibfnamefont {S.~A.}\ \bibnamefont
  {Maier}},\ }\href@noop {} {\emph {\bibinfo {title} {Plasmonics: fundamentals
  and applications}}}\ (\bibinfo  {publisher} {Springer Science \& Business
  Media},\ \bibinfo {year} {2007})\BibitemShut {NoStop}%
\bibitem [{\citenamefont {Li}\ \emph {et~al.}(2003)\citenamefont {Li},
  \citenamefont {Stockman},\ and\ \citenamefont {Bergman}}]{Li2003}%
  \BibitemOpen
  \bibfield  {author} {\bibinfo {author} {\bibfnamefont {K.}~\bibnamefont
  {Li}}, \bibinfo {author} {\bibfnamefont {M.~I.}\ \bibnamefont {Stockman}}, \
  and\ \bibinfo {author} {\bibfnamefont {D.~J.}\ \bibnamefont {Bergman}},\
  }\href@noop {} {\bibfield  {journal} {\bibinfo  {journal} {Physical review
  letters}\ }\textbf {\bibinfo {volume} {91}},\ \bibinfo {pages} {227402}
  (\bibinfo {year} {2003})}\BibitemShut {NoStop}%
\bibitem [{\citenamefont {Mayer}\ and\ \citenamefont
  {Hafner}(2011)}]{Mayer2011}%
  \BibitemOpen
  \bibfield  {author} {\bibinfo {author} {\bibfnamefont {K.~M.}\ \bibnamefont
  {Mayer}}\ and\ \bibinfo {author} {\bibfnamefont {J.~H.}\ \bibnamefont
  {Hafner}},\ }\href@noop {} {\bibfield  {journal} {\bibinfo  {journal}
  {Chemical reviews}\ }\textbf {\bibinfo {volume} {111}},\ \bibinfo {pages}
  {3828} (\bibinfo {year} {2011})}\BibitemShut {NoStop}%
\bibitem [{\citenamefont {Atwater}\ and\ \citenamefont
  {Polman}(2010)}]{Atwater2010}%
  \BibitemOpen
  \bibfield  {author} {\bibinfo {author} {\bibfnamefont {H.~A.}\ \bibnamefont
  {Atwater}}\ and\ \bibinfo {author} {\bibfnamefont {A.}~\bibnamefont
  {Polman}},\ }\href@noop {} {\bibfield  {journal} {\bibinfo  {journal} {Nature
  materials}\ }\textbf {\bibinfo {volume} {9}},\ \bibinfo {pages} {205}
  (\bibinfo {year} {2010})}\BibitemShut {NoStop}%
\bibitem [{\citenamefont {Kauranen}\ and\ \citenamefont
  {Zayats}(2012)}]{Kauranen2012}%
  \BibitemOpen
  \bibfield  {author} {\bibinfo {author} {\bibfnamefont {M.}~\bibnamefont
  {Kauranen}}\ and\ \bibinfo {author} {\bibfnamefont {A.~V.}\ \bibnamefont
  {Zayats}},\ }\href@noop {} {\bibfield  {journal} {\bibinfo  {journal} {Nature
  Photonics}\ }\textbf {\bibinfo {volume} {6}},\ \bibinfo {pages} {737}
  (\bibinfo {year} {2012})}\BibitemShut {NoStop}%
\bibitem [{\citenamefont {Le~Ru}\ and\ \citenamefont
  {Etchegoin}(2008)}]{Le2008}%
  \BibitemOpen
  \bibfield  {author} {\bibinfo {author} {\bibfnamefont {E.}~\bibnamefont
  {Le~Ru}}\ and\ \bibinfo {author} {\bibfnamefont {P.}~\bibnamefont
  {Etchegoin}},\ }\href@noop {} {\emph {\bibinfo {title} {Principles of
  Surface-Enhanced Raman Spectroscopy: and related plasmonic effects}}}\
  (\bibinfo  {publisher} {Elsevier},\ \bibinfo {year} {2008})\BibitemShut
  {NoStop}%
\bibitem [{\citenamefont {Bidault}\ \emph {et~al.}(2008)\citenamefont
  {Bidault}, \citenamefont {García~de Abajo},\ and\ \citenamefont
  {Polman}}]{Bidault2007}%
  \BibitemOpen
  \bibfield  {author} {\bibinfo {author} {\bibfnamefont {S.}~\bibnamefont
  {Bidault}}, \bibinfo {author} {\bibfnamefont {F.~J.}\ \bibnamefont
  {García~de Abajo}}, \ and\ \bibinfo {author} {\bibfnamefont
  {A.}~\bibnamefont {Polman}},\ }\href {\doibase 10.1021/ja711074n} {\bibfield
  {journal} {\bibinfo  {journal} {Journal of the American Chemical Society}\
  }\textbf {\bibinfo {volume} {130}},\ \bibinfo {pages} {2750} (\bibinfo {year}
  {2008})},\ \bibinfo {note} {pMID: 18266376}\BibitemShut {NoStop}%
\bibitem [{\citenamefont {Kneipp}\ \emph {et~al.}(2008)\citenamefont {Kneipp},
  \citenamefont {Li}, \citenamefont {Sherwood}, \citenamefont {Panne},
  \citenamefont {Kneipp}, \citenamefont {Stockman},\ and\ \citenamefont
  {Kneipp}}]{Kneipp2008}%
  \BibitemOpen
  \bibfield  {author} {\bibinfo {author} {\bibfnamefont {J.}~\bibnamefont
  {Kneipp}}, \bibinfo {author} {\bibfnamefont {X.}~\bibnamefont {Li}}, \bibinfo
  {author} {\bibfnamefont {M.}~\bibnamefont {Sherwood}}, \bibinfo {author}
  {\bibfnamefont {U.}~\bibnamefont {Panne}}, \bibinfo {author} {\bibfnamefont
  {H.}~\bibnamefont {Kneipp}}, \bibinfo {author} {\bibfnamefont {M.~I.}\
  \bibnamefont {Stockman}}, \ and\ \bibinfo {author} {\bibfnamefont
  {K.}~\bibnamefont {Kneipp}},\ }\href@noop {} {\bibfield  {journal} {\bibinfo
  {journal} {Analytical chemistry}\ }\textbf {\bibinfo {volume} {80}},\
  \bibinfo {pages} {4247} (\bibinfo {year} {2008})}\BibitemShut {NoStop}%
\bibitem [{\citenamefont {Ding}\ \emph {et~al.}(2010)\citenamefont {Ding},
  \citenamefont {Deng}, \citenamefont {Yan}, \citenamefont {Cabrini},
  \citenamefont {Zuckermann},\ and\ \citenamefont {Bokor}}]{Ding2010}%
  \BibitemOpen
  \bibfield  {author} {\bibinfo {author} {\bibfnamefont {B.}~\bibnamefont
  {Ding}}, \bibinfo {author} {\bibfnamefont {Z.}~\bibnamefont {Deng}}, \bibinfo
  {author} {\bibfnamefont {H.}~\bibnamefont {Yan}}, \bibinfo {author}
  {\bibfnamefont {S.}~\bibnamefont {Cabrini}}, \bibinfo {author} {\bibfnamefont
  {R.~N.}\ \bibnamefont {Zuckermann}}, \ and\ \bibinfo {author} {\bibfnamefont
  {J.}~\bibnamefont {Bokor}},\ }\href@noop {} {\bibfield  {journal} {\bibinfo
  {journal} {Journal of the American Chemical Society}\ }\textbf {\bibinfo
  {volume} {132}},\ \bibinfo {pages} {3248} (\bibinfo {year}
  {2010})}\BibitemShut {NoStop}%
\bibitem [{\citenamefont {Kravets}\ \emph {et~al.}(2010)\citenamefont
  {Kravets}, \citenamefont {Zoriniants}, \citenamefont {Burrows}, \citenamefont
  {Schedin}, \citenamefont {Casiraghi}, \citenamefont {Klar}, \citenamefont
  {Geim}, \citenamefont {Barnes},\ and\ \citenamefont
  {Grigorenko}}]{Kravets2010}%
  \BibitemOpen
  \bibfield  {author} {\bibinfo {author} {\bibfnamefont {V.}~\bibnamefont
  {Kravets}}, \bibinfo {author} {\bibfnamefont {G.}~\bibnamefont {Zoriniants}},
  \bibinfo {author} {\bibfnamefont {C.~P.}\ \bibnamefont {Burrows}}, \bibinfo
  {author} {\bibfnamefont {F.}~\bibnamefont {Schedin}}, \bibinfo {author}
  {\bibfnamefont {C.}~\bibnamefont {Casiraghi}}, \bibinfo {author}
  {\bibfnamefont {P.}~\bibnamefont {Klar}}, \bibinfo {author} {\bibfnamefont
  {A.}~\bibnamefont {Geim}}, \bibinfo {author} {\bibfnamefont {W.~L.}\
  \bibnamefont {Barnes}}, \ and\ \bibinfo {author} {\bibfnamefont
  {A.}~\bibnamefont {Grigorenko}},\ }\href@noop {} {\bibfield  {journal}
  {\bibinfo  {journal} {Physical review letters}\ }\textbf {\bibinfo {volume}
  {105}},\ \bibinfo {pages} {246806} (\bibinfo {year} {2010})}\BibitemShut
  {NoStop}%
\bibitem [{\citenamefont {H{\"o}ppener}\ \emph {et~al.}(2012)\citenamefont
  {H{\"o}ppener}, \citenamefont {Lapin}, \citenamefont {Bharadwaj},\ and\
  \citenamefont {Novotny}}]{Hoppener2012}%
  \BibitemOpen
  \bibfield  {author} {\bibinfo {author} {\bibfnamefont {C.}~\bibnamefont
  {H{\"o}ppener}}, \bibinfo {author} {\bibfnamefont {Z.~J.}\ \bibnamefont
  {Lapin}}, \bibinfo {author} {\bibfnamefont {P.}~\bibnamefont {Bharadwaj}}, \
  and\ \bibinfo {author} {\bibfnamefont {L.}~\bibnamefont {Novotny}},\
  }\href@noop {} {\bibfield  {journal} {\bibinfo  {journal} {Physical review
  letters}\ }\textbf {\bibinfo {volume} {109}},\ \bibinfo {pages} {017402}
  (\bibinfo {year} {2012})}\BibitemShut {NoStop}%
\bibitem [{\citenamefont {Coluccio}\ \emph {et~al.}(2015)\citenamefont
  {Coluccio}, \citenamefont {Gentile}, \citenamefont {Das}, \citenamefont
  {Nicastri}, \citenamefont {Perri}, \citenamefont {Candeloro}, \citenamefont
  {Perozziello}, \citenamefont {Zaccaria}, \citenamefont {Gongora},
  \citenamefont {Alrasheed} \emph {et~al.}}]{Coluccio2015}%
  \BibitemOpen
  \bibfield  {author} {\bibinfo {author} {\bibfnamefont {M.~L.}\ \bibnamefont
  {Coluccio}}, \bibinfo {author} {\bibfnamefont {F.}~\bibnamefont {Gentile}},
  \bibinfo {author} {\bibfnamefont {G.}~\bibnamefont {Das}}, \bibinfo {author}
  {\bibfnamefont {A.}~\bibnamefont {Nicastri}}, \bibinfo {author}
  {\bibfnamefont {A.~M.}\ \bibnamefont {Perri}}, \bibinfo {author}
  {\bibfnamefont {P.}~\bibnamefont {Candeloro}}, \bibinfo {author}
  {\bibfnamefont {G.}~\bibnamefont {Perozziello}}, \bibinfo {author}
  {\bibfnamefont {R.~P.}\ \bibnamefont {Zaccaria}}, \bibinfo {author}
  {\bibfnamefont {J.~S.~T.}\ \bibnamefont {Gongora}}, \bibinfo {author}
  {\bibfnamefont {S.}~\bibnamefont {Alrasheed}},  \emph {et~al.},\ }\href@noop
  {} {\bibfield  {journal} {\bibinfo  {journal} {Science advances}\ }\textbf
  {\bibinfo {volume} {1}},\ \bibinfo {pages} {e1500487} (\bibinfo {year}
  {2015})}\BibitemShut {NoStop}%
\bibitem [{\citenamefont {Laura{\'a}Coluccio}\ \emph
  {et~al.}(2016)\citenamefont {Laura{\'a}Coluccio}, \citenamefont {Fabrizio}
  \emph {et~al.}}]{Lauraacoluccio2016}%
  \BibitemOpen
  \bibfield  {author} {\bibinfo {author} {\bibfnamefont {M.}~\bibnamefont
  {Laura{\'a}Coluccio}}, \bibinfo {author} {\bibfnamefont {E.}~\bibnamefont
  {Fabrizio}},  \emph {et~al.},\ }\href@noop {} {\bibfield  {journal} {\bibinfo
   {journal} {RSC Advances}\ }\textbf {\bibinfo {volume} {6}},\ \bibinfo
  {pages} {107916} (\bibinfo {year} {2016})}\BibitemShut {NoStop}%
\bibitem [{\citenamefont {Heck}\ \emph {et~al.}(2017)\citenamefont {Heck},
  \citenamefont {Prinz}, \citenamefont {Dathe}, \citenamefont {Merk},
  \citenamefont {Stranik}, \citenamefont {Fritzsche}, \citenamefont {Kneipp},\
  and\ \citenamefont {Bald}}]{Heck2017}%
  \BibitemOpen
  \bibfield  {author} {\bibinfo {author} {\bibfnamefont {C.}~\bibnamefont
  {Heck}}, \bibinfo {author} {\bibfnamefont {J.}~\bibnamefont {Prinz}},
  \bibinfo {author} {\bibfnamefont {A.}~\bibnamefont {Dathe}}, \bibinfo
  {author} {\bibfnamefont {V.}~\bibnamefont {Merk}}, \bibinfo {author}
  {\bibfnamefont {O.}~\bibnamefont {Stranik}}, \bibinfo {author} {\bibfnamefont
  {W.}~\bibnamefont {Fritzsche}}, \bibinfo {author} {\bibfnamefont
  {J.}~\bibnamefont {Kneipp}}, \ and\ \bibinfo {author} {\bibfnamefont
  {I.}~\bibnamefont {Bald}},\ }\href@noop {} {\bibfield  {journal} {\bibinfo
  {journal} {ACS Photonics}\ } (\bibinfo {year} {2017})}\BibitemShut {NoStop}%
\bibitem [{\citenamefont {Lloyd}\ \emph {et~al.}(2017)\citenamefont {Lloyd},
  \citenamefont {Ng}, \citenamefont {Liu}, \citenamefont {Zhu}, \citenamefont
  {Chao}, \citenamefont {Coenen}, \citenamefont {Etheridge}, \citenamefont
  {G{\'o}mez},\ and\ \citenamefont {Bach}}]{Lloyd2017}%
  \BibitemOpen
  \bibfield  {author} {\bibinfo {author} {\bibfnamefont {J.~A.}\ \bibnamefont
  {Lloyd}}, \bibinfo {author} {\bibfnamefont {S.~H.}\ \bibnamefont {Ng}},
  \bibinfo {author} {\bibfnamefont {A.~C.}\ \bibnamefont {Liu}}, \bibinfo
  {author} {\bibfnamefont {Y.}~\bibnamefont {Zhu}}, \bibinfo {author}
  {\bibfnamefont {W.}~\bibnamefont {Chao}}, \bibinfo {author} {\bibfnamefont
  {T.}~\bibnamefont {Coenen}}, \bibinfo {author} {\bibfnamefont
  {J.}~\bibnamefont {Etheridge}}, \bibinfo {author} {\bibfnamefont {D.~E.}\
  \bibnamefont {G{\'o}mez}}, \ and\ \bibinfo {author} {\bibfnamefont
  {U.}~\bibnamefont {Bach}},\ }\href@noop {} {\bibfield  {journal} {\bibinfo
  {journal} {ACS nano}\ }\textbf {\bibinfo {volume} {11}},\ \bibinfo {pages}
  {1604} (\bibinfo {year} {2017})}\BibitemShut {NoStop}%
\bibitem [{\citenamefont {Li}\ \emph {et~al.}(2006)\citenamefont {Li},
  \citenamefont {Yang},\ and\ \citenamefont {Xu}}]{Li2006}%
  \BibitemOpen
  \bibfield  {author} {\bibinfo {author} {\bibfnamefont {Z.}~\bibnamefont
  {Li}}, \bibinfo {author} {\bibfnamefont {Z.}~\bibnamefont {Yang}}, \ and\
  \bibinfo {author} {\bibfnamefont {H.}~\bibnamefont {Xu}},\ }\href@noop {}
  {\bibfield  {journal} {\bibinfo  {journal} {Physical review letters}\
  }\textbf {\bibinfo {volume} {97}},\ \bibinfo {pages} {079701} (\bibinfo
  {year} {2006})}\BibitemShut {NoStop}%
\bibitem [{\citenamefont {Dai}\ \emph {et~al.}(2008)\citenamefont {Dai},
  \citenamefont {{\v{C}}ajko}, \citenamefont {Tsukerman},\ and\ \citenamefont
  {Stockman}}]{Dai2008}%
  \BibitemOpen
  \bibfield  {author} {\bibinfo {author} {\bibfnamefont {J.}~\bibnamefont
  {Dai}}, \bibinfo {author} {\bibfnamefont {F.}~\bibnamefont {{\v{C}}ajko}},
  \bibinfo {author} {\bibfnamefont {I.}~\bibnamefont {Tsukerman}}, \ and\
  \bibinfo {author} {\bibfnamefont {M.~I.}\ \bibnamefont {Stockman}},\
  }\href@noop {} {\bibfield  {journal} {\bibinfo  {journal} {Physical Review
  B}\ }\textbf {\bibinfo {volume} {77}},\ \bibinfo {pages} {115419} (\bibinfo
  {year} {2008})}\BibitemShut {NoStop}%
\bibitem [{\citenamefont {McMahon}\ \emph {et~al.}(2009)\citenamefont
  {McMahon}, \citenamefont {Gray},\ and\ \citenamefont {Schatz}}]{Mcmahon2009}%
  \BibitemOpen
  \bibfield  {author} {\bibinfo {author} {\bibfnamefont {J.~M.}\ \bibnamefont
  {McMahon}}, \bibinfo {author} {\bibfnamefont {S.~K.}\ \bibnamefont {Gray}}, \
  and\ \bibinfo {author} {\bibfnamefont {G.~C.}\ \bibnamefont {Schatz}},\
  }\href@noop {} {\bibfield  {journal} {\bibinfo  {journal} {Physical review
  letters}\ }\textbf {\bibinfo {volume} {103}},\ \bibinfo {pages} {097403}
  (\bibinfo {year} {2009})}\BibitemShut {NoStop}%
\bibitem [{\citenamefont {David}\ and\ \citenamefont {García~de
  Abajo}(2011)}]{David2011}%
  \BibitemOpen
  \bibfield  {author} {\bibinfo {author} {\bibfnamefont {C.}~\bibnamefont
  {David}}\ and\ \bibinfo {author} {\bibfnamefont {F.~J.}\ \bibnamefont
  {García~de Abajo}},\ }\href@noop {} {\bibfield  {journal} {\bibinfo
  {journal} {The Journal of Physical Chemistry C}\ }\textbf {\bibinfo {volume}
  {115}},\ \bibinfo {pages} {19470} (\bibinfo {year} {2011})}\BibitemShut
  {NoStop}%
\bibitem [{\citenamefont {Toscano}\ \emph {et~al.}(2012)\citenamefont
  {Toscano}, \citenamefont {Raza}, \citenamefont {Jauho}, \citenamefont
  {Mortensen},\ and\ \citenamefont {Wubs}}]{Toscano2012}%
  \BibitemOpen
  \bibfield  {author} {\bibinfo {author} {\bibfnamefont {G.}~\bibnamefont
  {Toscano}}, \bibinfo {author} {\bibfnamefont {S.}~\bibnamefont {Raza}},
  \bibinfo {author} {\bibfnamefont {A.-P.}\ \bibnamefont {Jauho}}, \bibinfo
  {author} {\bibfnamefont {N.~A.}\ \bibnamefont {Mortensen}}, \ and\ \bibinfo
  {author} {\bibfnamefont {M.}~\bibnamefont {Wubs}},\ }\href@noop {} {\bibfield
   {journal} {\bibinfo  {journal} {Optics express}\ }\textbf {\bibinfo {volume}
  {20}},\ \bibinfo {pages} {4176} (\bibinfo {year} {2012})}\BibitemShut
  {NoStop}%
\bibitem [{\citenamefont {Luo}\ \emph {et~al.}(2013)\citenamefont {Luo},
  \citenamefont {Fernandez-Dominguez}, \citenamefont {Wiener}, \citenamefont
  {Maier},\ and\ \citenamefont {Pendry}}]{Luo2013}%
  \BibitemOpen
  \bibfield  {author} {\bibinfo {author} {\bibfnamefont {Y.}~\bibnamefont
  {Luo}}, \bibinfo {author} {\bibfnamefont {A.}~\bibnamefont
  {Fernandez-Dominguez}}, \bibinfo {author} {\bibfnamefont {A.}~\bibnamefont
  {Wiener}}, \bibinfo {author} {\bibfnamefont {S.~A.}\ \bibnamefont {Maier}}, \
  and\ \bibinfo {author} {\bibfnamefont {J.}~\bibnamefont {Pendry}},\
  }\href@noop {} {\bibfield  {journal} {\bibinfo  {journal} {Physical review
  letters}\ }\textbf {\bibinfo {volume} {111}},\ \bibinfo {pages} {093901}
  (\bibinfo {year} {2013})}\BibitemShut {NoStop}%
\bibitem [{\citenamefont {Teperik}\ \emph {et~al.}(2013)\citenamefont
  {Teperik}, \citenamefont {Nordlander}, \citenamefont {Aizpurua},\ and\
  \citenamefont {Borisov}}]{Teperik2013}%
  \BibitemOpen
  \bibfield  {author} {\bibinfo {author} {\bibfnamefont {T.~V.}\ \bibnamefont
  {Teperik}}, \bibinfo {author} {\bibfnamefont {P.}~\bibnamefont {Nordlander}},
  \bibinfo {author} {\bibfnamefont {J.}~\bibnamefont {Aizpurua}}, \ and\
  \bibinfo {author} {\bibfnamefont {A.~G.}\ \bibnamefont {Borisov}},\
  }\href@noop {} {\bibfield  {journal} {\bibinfo  {journal} {Physical review
  letters}\ }\textbf {\bibinfo {volume} {110}},\ \bibinfo {pages} {263901}
  (\bibinfo {year} {2013})}\BibitemShut {NoStop}%
\bibitem [{\citenamefont {Mortensen}\ \emph {et~al.}(2014)\citenamefont
  {Mortensen}, \citenamefont {Raza}, \citenamefont {Wubs}, \citenamefont
  {S{\o}ndergaard},\ and\ \citenamefont {Bozhevolnyi}}]{Mortensen2014}%
  \BibitemOpen
  \bibfield  {author} {\bibinfo {author} {\bibfnamefont {N.~A.}\ \bibnamefont
  {Mortensen}}, \bibinfo {author} {\bibfnamefont {S.}~\bibnamefont {Raza}},
  \bibinfo {author} {\bibfnamefont {M.}~\bibnamefont {Wubs}}, \bibinfo {author}
  {\bibfnamefont {T.}~\bibnamefont {S{\o}ndergaard}}, \ and\ \bibinfo {author}
  {\bibfnamefont {S.~I.}\ \bibnamefont {Bozhevolnyi}},\ }\href@noop {}
  {\bibfield  {journal} {\bibinfo  {journal} {Nature communications}\ }\textbf
  {\bibinfo {volume} {5}} (\bibinfo {year} {2014})}\BibitemShut {NoStop}%
\bibitem [{\citenamefont {Schnitzer}\ \emph {et~al.}(2016)\citenamefont
  {Schnitzer}, \citenamefont {Giannini}, \citenamefont {Maier},\ and\
  \citenamefont {Craster}}]{Schnitzer2016}%
  \BibitemOpen
  \bibfield  {author} {\bibinfo {author} {\bibfnamefont {O.}~\bibnamefont
  {Schnitzer}}, \bibinfo {author} {\bibfnamefont {V.}~\bibnamefont {Giannini}},
  \bibinfo {author} {\bibfnamefont {S.~A.}\ \bibnamefont {Maier}}, \ and\
  \bibinfo {author} {\bibfnamefont {R.~V.}\ \bibnamefont {Craster}},\ }in\
  \href@noop {} {\emph {\bibinfo {booktitle} {Proc. R. Soc. A}}},\ Vol.\
  \bibinfo {volume} {472}\ (\bibinfo {organization} {The Royal Society},\
  \bibinfo {year} {2016})\ p.\ \bibinfo {pages} {20160258}\BibitemShut
  {NoStop}%
\bibitem [{\citenamefont {Toscano}\ \emph {et~al.}(2015)\citenamefont
  {Toscano}, \citenamefont {Straubel}, \citenamefont {Kwiatkowski},
  \citenamefont {Rockstuhl}, \citenamefont {Evers}, \citenamefont {Xu},
  \citenamefont {Mortensen},\ and\ \citenamefont {Wubs}}]{Toscano2015}%
  \BibitemOpen
  \bibfield  {author} {\bibinfo {author} {\bibfnamefont {G.}~\bibnamefont
  {Toscano}}, \bibinfo {author} {\bibfnamefont {J.}~\bibnamefont {Straubel}},
  \bibinfo {author} {\bibfnamefont {A.}~\bibnamefont {Kwiatkowski}}, \bibinfo
  {author} {\bibfnamefont {C.}~\bibnamefont {Rockstuhl}}, \bibinfo {author}
  {\bibfnamefont {F.}~\bibnamefont {Evers}}, \bibinfo {author} {\bibfnamefont
  {H.}~\bibnamefont {Xu}}, \bibinfo {author} {\bibfnamefont {N.~A.}\
  \bibnamefont {Mortensen}}, \ and\ \bibinfo {author} {\bibfnamefont
  {M.}~\bibnamefont {Wubs}},\ }\href@noop {} {\bibfield  {journal} {\bibinfo
  {journal} {Nature communications}\ }\textbf {\bibinfo {volume} {6}} (\bibinfo
  {year} {2015})}\BibitemShut {NoStop}%
\bibitem [{\citenamefont {Zhang}\ \emph {et~al.}(2014)\citenamefont {Zhang},
  \citenamefont {Feist}, \citenamefont {Rubio}, \citenamefont
  {Garc{\'\i}a-Gonz{\'a}lez},\ and\ \citenamefont
  {Garc{\'\i}a-Vidal}}]{Zhang2014}%
  \BibitemOpen
  \bibfield  {author} {\bibinfo {author} {\bibfnamefont {P.}~\bibnamefont
  {Zhang}}, \bibinfo {author} {\bibfnamefont {J.}~\bibnamefont {Feist}},
  \bibinfo {author} {\bibfnamefont {A.}~\bibnamefont {Rubio}}, \bibinfo
  {author} {\bibfnamefont {P.}~\bibnamefont {Garc{\'\i}a-Gonz{\'a}lez}}, \ and\
  \bibinfo {author} {\bibfnamefont {F.}~\bibnamefont {Garc{\'\i}a-Vidal}},\
  }\href@noop {} {\bibfield  {journal} {\bibinfo  {journal} {Physical Review
  B}\ }\textbf {\bibinfo {volume} {90}},\ \bibinfo {pages} {161407} (\bibinfo
  {year} {2014})}\BibitemShut {NoStop}%
\bibitem [{\citenamefont {Varas}\ \emph {et~al.}(2016)\citenamefont {Varas},
  \citenamefont {Garc{\'\i}a-Gonz{\'a}lez}, \citenamefont {Feist},
  \citenamefont {Garc{\'\i}a-Vidal},\ and\ \citenamefont {Rubio}}]{Varas2016}%
  \BibitemOpen
  \bibfield  {author} {\bibinfo {author} {\bibfnamefont {A.}~\bibnamefont
  {Varas}}, \bibinfo {author} {\bibfnamefont {P.}~\bibnamefont
  {Garc{\'\i}a-Gonz{\'a}lez}}, \bibinfo {author} {\bibfnamefont
  {J.}~\bibnamefont {Feist}}, \bibinfo {author} {\bibfnamefont
  {F.}~\bibnamefont {Garc{\'\i}a-Vidal}}, \ and\ \bibinfo {author}
  {\bibfnamefont {A.}~\bibnamefont {Rubio}},\ }\href@noop {} {\bibfield
  {journal} {\bibinfo  {journal} {Nanophotonics}\ }\textbf {\bibinfo {volume}
  {5}},\ \bibinfo {pages} {409} (\bibinfo {year} {2016})}\BibitemShut {NoStop}%
\bibitem [{\citenamefont {Fitzgerald}\ \emph {et~al.}(2016)\citenamefont
  {Fitzgerald}, \citenamefont {Narang}, \citenamefont {Craster}, \citenamefont
  {Maier},\ and\ \citenamefont {Giannini}}]{Fitzgerald2016}%
  \BibitemOpen
  \bibfield  {author} {\bibinfo {author} {\bibfnamefont {J.~M.}\ \bibnamefont
  {Fitzgerald}}, \bibinfo {author} {\bibfnamefont {P.}~\bibnamefont {Narang}},
  \bibinfo {author} {\bibfnamefont {R.~V.}\ \bibnamefont {Craster}}, \bibinfo
  {author} {\bibfnamefont {S.~A.}\ \bibnamefont {Maier}}, \ and\ \bibinfo
  {author} {\bibfnamefont {V.}~\bibnamefont {Giannini}},\ }\href {\doibase
  10.1109/JPROC.2016.2584860} {\bibfield  {journal} {\bibinfo  {journal}
  {Proceedings of the IEEE}\ }\textbf {\bibinfo {volume} {104}},\ \bibinfo
  {pages} {2307} (\bibinfo {year} {2016})}\BibitemShut {NoStop}%
\bibitem [{\citenamefont {Bursi}\ \emph {et~al.}(2014)\citenamefont {Bursi},
  \citenamefont {Calzolari}, \citenamefont {Corni},\ and\ \citenamefont
  {Molinari}}]{Bursi2014}%
  \BibitemOpen
  \bibfield  {author} {\bibinfo {author} {\bibfnamefont {L.}~\bibnamefont
  {Bursi}}, \bibinfo {author} {\bibfnamefont {A.}~\bibnamefont {Calzolari}},
  \bibinfo {author} {\bibfnamefont {S.}~\bibnamefont {Corni}}, \ and\ \bibinfo
  {author} {\bibfnamefont {E.}~\bibnamefont {Molinari}},\ }\href@noop {}
  {\bibfield  {journal} {\bibinfo  {journal} {ACS Photonics}\ }\textbf
  {\bibinfo {volume} {1}},\ \bibinfo {pages} {1049} (\bibinfo {year}
  {2014})}\BibitemShut {NoStop}%
\bibitem [{\citenamefont {Fitzgerald}\ \emph {et~al.}(2017)\citenamefont
  {Fitzgerald}, \citenamefont {Azadi},\ and\ \citenamefont
  {Giannini}}]{Fitzgerald2017}%
  \BibitemOpen
  \bibfield  {author} {\bibinfo {author} {\bibfnamefont {J.~M.}\ \bibnamefont
  {Fitzgerald}}, \bibinfo {author} {\bibfnamefont {S.}~\bibnamefont {Azadi}}, \
  and\ \bibinfo {author} {\bibfnamefont {V.}~\bibnamefont {Giannini}},\
  }\href@noop {} {\bibfield  {journal} {\bibinfo  {journal} {Physical Review
  B}\ }\textbf {\bibinfo {volume} {95}},\ \bibinfo {pages} {235414} (\bibinfo
  {year} {2017})}\BibitemShut {NoStop}%
\bibitem [{\citenamefont {Fitzgerald}\ and\ \citenamefont
  {Giannini}(2017)}]{Fitzgerald2017perspective}%
  \BibitemOpen
  \bibfield  {author} {\bibinfo {author} {\bibfnamefont {J.~M.}\ \bibnamefont
  {Fitzgerald}}\ and\ \bibinfo {author} {\bibfnamefont {V.}~\bibnamefont
  {Giannini}},\ }\href {http://stacks.iop.org/2040-8986/19/i=6/a=060401}
  {\bibfield  {journal} {\bibinfo  {journal} {Journal of Optics}\ }\textbf
  {\bibinfo {volume} {19}},\ \bibinfo {pages} {060401} (\bibinfo {year}
  {2017})}\BibitemShut {NoStop}%
\bibitem [{\citenamefont {Kreibig}\ and\ \citenamefont
  {Vollmer}(2013)}]{Kreibig2013}%
  \BibitemOpen
  \bibfield  {author} {\bibinfo {author} {\bibfnamefont {U.}~\bibnamefont
  {Kreibig}}\ and\ \bibinfo {author} {\bibfnamefont {M.}~\bibnamefont
  {Vollmer}},\ }\href@noop {} {\emph {\bibinfo {title} {Optical properties of
  metal clusters}}},\ Vol.~\bibinfo {volume} {25}\ (\bibinfo  {publisher}
  {Springer Science \& Business Media},\ \bibinfo {year} {2013})\BibitemShut
  {NoStop}%
\bibitem [{\citenamefont {Donati}\ \emph {et~al.}(2017)\citenamefont {Donati},
  \citenamefont {Lingerfelt}, \citenamefont {Aikens},\ and\ \citenamefont
  {Li}}]{Donati2017}%
  \BibitemOpen
  \bibfield  {author} {\bibinfo {author} {\bibfnamefont {G.}~\bibnamefont
  {Donati}}, \bibinfo {author} {\bibfnamefont {D.~B.}\ \bibnamefont
  {Lingerfelt}}, \bibinfo {author} {\bibfnamefont {C.~M.}\ \bibnamefont
  {Aikens}}, \ and\ \bibinfo {author} {\bibfnamefont {X.}~\bibnamefont {Li}},\
  }\href@noop {} {\bibfield  {journal} {\bibinfo  {journal} {The Journal of
  Physical Chemistry C}\ } (\bibinfo {year} {2017})}\BibitemShut {NoStop}%
\bibitem [{\citenamefont {Kreibig}(1978)}]{Kreibig1978}%
  \BibitemOpen
  \bibfield  {author} {\bibinfo {author} {\bibfnamefont {U.}~\bibnamefont
  {Kreibig}},\ }\href@noop {} {\bibfield  {journal} {\bibinfo  {journal} {Solid
  State Communications}\ }\textbf {\bibinfo {volume} {28}},\ \bibinfo {pages}
  {767} (\bibinfo {year} {1978})}\BibitemShut {NoStop}%
\bibitem [{\citenamefont {Pinchuk}\ and\ \citenamefont
  {Kreibig}(2003)}]{Pinchuk2003}%
  \BibitemOpen
  \bibfield  {author} {\bibinfo {author} {\bibfnamefont {A.}~\bibnamefont
  {Pinchuk}}\ and\ \bibinfo {author} {\bibfnamefont {U.}~\bibnamefont
  {Kreibig}},\ }\href@noop {} {\bibfield  {journal} {\bibinfo  {journal} {New
  Journal of Physics}\ }\textbf {\bibinfo {volume} {5}},\ \bibinfo {pages}
  {151} (\bibinfo {year} {2003})}\BibitemShut {NoStop}%
\bibitem [{\citenamefont {Molina}\ \emph {et~al.}(2002)\citenamefont {Molina},
  \citenamefont {Weinmann},\ and\ \citenamefont {Jalabert}}]{Molina2002}%
  \BibitemOpen
  \bibfield  {author} {\bibinfo {author} {\bibfnamefont {R.~A.}\ \bibnamefont
  {Molina}}, \bibinfo {author} {\bibfnamefont {D.}~\bibnamefont {Weinmann}}, \
  and\ \bibinfo {author} {\bibfnamefont {R.~A.}\ \bibnamefont {Jalabert}},\
  }\href@noop {} {\bibfield  {journal} {\bibinfo  {journal} {Physical Review
  B}\ }\textbf {\bibinfo {volume} {65}},\ \bibinfo {pages} {155427} (\bibinfo
  {year} {2002})}\BibitemShut {NoStop}%
\bibitem [{\citenamefont {Lerm{\'e}}\ \emph {et~al.}(2010)\citenamefont
  {Lerm{\'e}}, \citenamefont {Baida}, \citenamefont {Bonnet}, \citenamefont
  {Broyer}, \citenamefont {Cottancin}, \citenamefont {Crut}, \citenamefont
  {Maioli}, \citenamefont {Del~Fatti}, \citenamefont {Vall{\'e}e},\ and\
  \citenamefont {Pellarin}}]{Lerme2010}%
  \BibitemOpen
  \bibfield  {author} {\bibinfo {author} {\bibfnamefont {J.}~\bibnamefont
  {Lerm{\'e}}}, \bibinfo {author} {\bibfnamefont {H.}~\bibnamefont {Baida}},
  \bibinfo {author} {\bibfnamefont {C.}~\bibnamefont {Bonnet}}, \bibinfo
  {author} {\bibfnamefont {M.}~\bibnamefont {Broyer}}, \bibinfo {author}
  {\bibfnamefont {E.}~\bibnamefont {Cottancin}}, \bibinfo {author}
  {\bibfnamefont {A.}~\bibnamefont {Crut}}, \bibinfo {author} {\bibfnamefont
  {P.}~\bibnamefont {Maioli}}, \bibinfo {author} {\bibfnamefont
  {N.}~\bibnamefont {Del~Fatti}}, \bibinfo {author} {\bibfnamefont
  {F.}~\bibnamefont {Vall{\'e}e}}, \ and\ \bibinfo {author} {\bibfnamefont
  {M.}~\bibnamefont {Pellarin}},\ }\href@noop {} {\bibfield  {journal}
  {\bibinfo  {journal} {The Journal of Physical Chemistry Letters}\ }\textbf
  {\bibinfo {volume} {1}},\ \bibinfo {pages} {2922} (\bibinfo {year}
  {2010})}\BibitemShut {NoStop}%
\bibitem [{\citenamefont {Quinten}(1996)}]{Quinten1996}%
  \BibitemOpen
  \bibfield  {author} {\bibinfo {author} {\bibfnamefont {M.}~\bibnamefont
  {Quinten}},\ }\href@noop {} {\bibfield  {journal} {\bibinfo  {journal}
  {Zeitschrift f{\"u}r Physik B Condensed Matter}\ }\textbf {\bibinfo {volume}
  {101}},\ \bibinfo {pages} {211} (\bibinfo {year} {1996})}\BibitemShut
  {NoStop}%
\bibitem [{\citenamefont {Moroz}(2008)}]{Moroz2008}%
  \BibitemOpen
  \bibfield  {author} {\bibinfo {author} {\bibfnamefont {A.}~\bibnamefont
  {Moroz}},\ }\href@noop {} {\bibfield  {journal} {\bibinfo  {journal} {The
  Journal of Physical Chemistry C}\ }\textbf {\bibinfo {volume} {112}},\
  \bibinfo {pages} {10641} (\bibinfo {year} {2008})}\BibitemShut {NoStop}%
\bibitem [{\citenamefont {Caldwell}\ \emph {et~al.}(2015)\citenamefont
  {Caldwell}, \citenamefont {Lindsay}, \citenamefont {Giannini}, \citenamefont
  {Vurgaftman}, \citenamefont {Reinecke}, \citenamefont {Maier},\ and\
  \citenamefont {Glembocki}}]{Caldwell2015}%
  \BibitemOpen
  \bibfield  {author} {\bibinfo {author} {\bibfnamefont {J.~D.}\ \bibnamefont
  {Caldwell}}, \bibinfo {author} {\bibfnamefont {L.}~\bibnamefont {Lindsay}},
  \bibinfo {author} {\bibfnamefont {V.}~\bibnamefont {Giannini}}, \bibinfo
  {author} {\bibfnamefont {I.}~\bibnamefont {Vurgaftman}}, \bibinfo {author}
  {\bibfnamefont {T.~L.}\ \bibnamefont {Reinecke}}, \bibinfo {author}
  {\bibfnamefont {S.~A.}\ \bibnamefont {Maier}}, \ and\ \bibinfo {author}
  {\bibfnamefont {O.~J.}\ \bibnamefont {Glembocki}},\ }\href@noop {} {\bibfield
   {journal} {\bibinfo  {journal} {Nanophotonics}\ }\textbf {\bibinfo {volume}
  {4}},\ \bibinfo {pages} {44} (\bibinfo {year} {2015})}\BibitemShut {NoStop}%
\bibitem [{\citenamefont {De~Abajo}\ and\ \citenamefont
  {Howie}(2002)}]{Abajo2002}%
  \BibitemOpen
  \bibfield  {author} {\bibinfo {author} {\bibfnamefont {F.~G.}\ \bibnamefont
  {De~Abajo}}\ and\ \bibinfo {author} {\bibfnamefont {A.}~\bibnamefont
  {Howie}},\ }\href@noop {} {\bibfield  {journal} {\bibinfo  {journal}
  {Physical Review B}\ }\textbf {\bibinfo {volume} {65}},\ \bibinfo {pages}
  {115418} (\bibinfo {year} {2002})}\BibitemShut {NoStop}%
\bibitem [{\citenamefont {Hohenester}\ and\ \citenamefont
  {Tr{\"u}gler}(2012)}]{Hohenester2012}%
  \BibitemOpen
  \bibfield  {author} {\bibinfo {author} {\bibfnamefont {U.}~\bibnamefont
  {Hohenester}}\ and\ \bibinfo {author} {\bibfnamefont {A.}~\bibnamefont
  {Tr{\"u}gler}},\ }\href@noop {} {\bibfield  {journal} {\bibinfo  {journal}
  {Computer Physics Communications}\ }\textbf {\bibinfo {volume} {183}},\
  \bibinfo {pages} {370} (\bibinfo {year} {2012})}\BibitemShut {NoStop}%
\bibitem [{\citenamefont {Waxenegger}\ \emph {et~al.}(2015)\citenamefont
  {Waxenegger}, \citenamefont {Tr{\"u}gler},\ and\ \citenamefont
  {Hohenester}}]{Waxenegger2015}%
  \BibitemOpen
  \bibfield  {author} {\bibinfo {author} {\bibfnamefont {J.}~\bibnamefont
  {Waxenegger}}, \bibinfo {author} {\bibfnamefont {A.}~\bibnamefont
  {Tr{\"u}gler}}, \ and\ \bibinfo {author} {\bibfnamefont {U.}~\bibnamefont
  {Hohenester}},\ }\href@noop {} {\bibfield  {journal} {\bibinfo  {journal}
  {Computer Physics Communications}\ }\textbf {\bibinfo {volume} {193}},\
  \bibinfo {pages} {138} (\bibinfo {year} {2015})}\BibitemShut {NoStop}%
\bibitem [{\citenamefont {Sun}\ \emph {et~al.}(2011)\citenamefont {Sun},
  \citenamefont {Khurgin},\ and\ \citenamefont {Bratkovsky}}]{Sun2011}%
  \BibitemOpen
  \bibfield  {author} {\bibinfo {author} {\bibfnamefont {G.}~\bibnamefont
  {Sun}}, \bibinfo {author} {\bibfnamefont {J.~B.}\ \bibnamefont {Khurgin}}, \
  and\ \bibinfo {author} {\bibfnamefont {A.}~\bibnamefont {Bratkovsky}},\
  }\href@noop {} {\bibfield  {journal} {\bibinfo  {journal} {Physical Review
  B}\ }\textbf {\bibinfo {volume} {84}},\ \bibinfo {pages} {045415} (\bibinfo
  {year} {2011})}\BibitemShut {NoStop}%
\bibitem [{\citenamefont {Esteban}\ \emph {et~al.}(2012)\citenamefont
  {Esteban}, \citenamefont {Borisov}, \citenamefont {Nordlander},\ and\
  \citenamefont {Aizpurua}}]{Esteban2012}%
  \BibitemOpen
  \bibfield  {author} {\bibinfo {author} {\bibfnamefont {R.}~\bibnamefont
  {Esteban}}, \bibinfo {author} {\bibfnamefont {A.~G.}\ \bibnamefont
  {Borisov}}, \bibinfo {author} {\bibfnamefont {P.}~\bibnamefont {Nordlander}},
  \ and\ \bibinfo {author} {\bibfnamefont {J.}~\bibnamefont {Aizpurua}},\
  }\href@noop {} {\bibfield  {journal} {\bibinfo  {journal} {Nature
  communications}\ }\textbf {\bibinfo {volume} {3}},\ \bibinfo {pages} {825}
  (\bibinfo {year} {2012})}\BibitemShut {NoStop}%
\bibitem [{\citenamefont {Pellegrini}\ \emph {et~al.}(2007)\citenamefont
  {Pellegrini}, \citenamefont {Mattei}, \citenamefont {Bello},\ and\
  \citenamefont {Mazzoldi}}]{Pellegrini2007}%
  \BibitemOpen
  \bibfield  {author} {\bibinfo {author} {\bibfnamefont {G.}~\bibnamefont
  {Pellegrini}}, \bibinfo {author} {\bibfnamefont {G.}~\bibnamefont {Mattei}},
  \bibinfo {author} {\bibfnamefont {V.}~\bibnamefont {Bello}}, \ and\ \bibinfo
  {author} {\bibfnamefont {P.}~\bibnamefont {Mazzoldi}},\ }\href@noop {}
  {\bibfield  {journal} {\bibinfo  {journal} {Materials Science and
  Engineering: C}\ }\textbf {\bibinfo {volume} {27}},\ \bibinfo {pages} {1347}
  (\bibinfo {year} {2007})}\BibitemShut {NoStop}%
\bibitem [{\citenamefont {Xia}\ \emph {et~al.}(2009)\citenamefont {Xia},
  \citenamefont {Yin},\ and\ \citenamefont {Kresin}}]{Xia2009}%
  \BibitemOpen
  \bibfield  {author} {\bibinfo {author} {\bibfnamefont {C.}~\bibnamefont
  {Xia}}, \bibinfo {author} {\bibfnamefont {C.}~\bibnamefont {Yin}}, \ and\
  \bibinfo {author} {\bibfnamefont {V.~V.}\ \bibnamefont {Kresin}},\
  }\href@noop {} {\bibfield  {journal} {\bibinfo  {journal} {Physical review
  letters}\ }\textbf {\bibinfo {volume} {102}},\ \bibinfo {pages} {156802}
  (\bibinfo {year} {2009})}\BibitemShut {NoStop}%
\bibitem [{\citenamefont {Monreal}\ \emph {et~al.}(2013)\citenamefont
  {Monreal}, \citenamefont {Antosiewicz},\ and\ \citenamefont
  {Apell}}]{Monreal2013}%
  \BibitemOpen
  \bibfield  {author} {\bibinfo {author} {\bibfnamefont {R.~C.}\ \bibnamefont
  {Monreal}}, \bibinfo {author} {\bibfnamefont {T.~J.}\ \bibnamefont
  {Antosiewicz}}, \ and\ \bibinfo {author} {\bibfnamefont {S.~P.}\ \bibnamefont
  {Apell}},\ }\href@noop {} {\bibfield  {journal} {\bibinfo  {journal} {New
  Journal of Physics}\ }\textbf {\bibinfo {volume} {15}},\ \bibinfo {pages}
  {083044} (\bibinfo {year} {2013})}\BibitemShut {NoStop}%
\bibitem [{\citenamefont {Reiners}\ \emph {et~al.}(1995)\citenamefont
  {Reiners}, \citenamefont {Ellert}, \citenamefont {Schmidt},\ and\
  \citenamefont {Haberland}}]{Reiners1995}%
  \BibitemOpen
  \bibfield  {author} {\bibinfo {author} {\bibfnamefont {T.}~\bibnamefont
  {Reiners}}, \bibinfo {author} {\bibfnamefont {C.}~\bibnamefont {Ellert}},
  \bibinfo {author} {\bibfnamefont {M.}~\bibnamefont {Schmidt}}, \ and\
  \bibinfo {author} {\bibfnamefont {H.}~\bibnamefont {Haberland}},\ }\href@noop
  {} {\bibfield  {journal} {\bibinfo  {journal} {Physical review letters}\
  }\textbf {\bibinfo {volume} {74}},\ \bibinfo {pages} {1558} (\bibinfo {year}
  {1995})}\BibitemShut {NoStop}%
\bibitem [{\citenamefont {Bohren}\ and\ \citenamefont
  {Huffman}(2008)}]{Bohren2008}%
  \BibitemOpen
  \bibfield  {author} {\bibinfo {author} {\bibfnamefont {C.~F.}\ \bibnamefont
  {Bohren}}\ and\ \bibinfo {author} {\bibfnamefont {D.~R.}\ \bibnamefont
  {Huffman}},\ }\href@noop {} {\emph {\bibinfo {title} {Absorption and
  scattering of light by small particles}}}\ (\bibinfo  {publisher} {John Wiley
  \& Sons},\ \bibinfo {year} {2008})\BibitemShut {NoStop}%
\bibitem [{\citenamefont {Yan}(2015)}]{Yan2015}%
  \BibitemOpen
  \bibfield  {author} {\bibinfo {author} {\bibfnamefont {W.}~\bibnamefont
  {Yan}},\ }\href@noop {} {\bibfield  {journal} {\bibinfo  {journal} {Physical
  Review B}\ }\textbf {\bibinfo {volume} {91}},\ \bibinfo {pages} {115416}
  (\bibinfo {year} {2015})}\BibitemShut {NoStop}%
\bibitem [{\citenamefont {Jurga}\ \emph {et~al.}(2017)\citenamefont {Jurga},
  \citenamefont {D'Agostino}, \citenamefont {Della~Sala},\ and\ \citenamefont
  {Ciraci}}]{Jurga2017}%
  \BibitemOpen
  \bibfield  {author} {\bibinfo {author} {\bibfnamefont {R.}~\bibnamefont
  {Jurga}}, \bibinfo {author} {\bibfnamefont {S.}~\bibnamefont {D'Agostino}},
  \bibinfo {author} {\bibfnamefont {F.}~\bibnamefont {Della~Sala}}, \ and\
  \bibinfo {author} {\bibfnamefont {C.}~\bibnamefont {Ciraci}},\ }\href@noop {}
  {\bibfield  {journal} {\bibinfo  {journal} {The Journal of Physical Chemistry
  C}\ } (\bibinfo {year} {2017})}\BibitemShut {NoStop}%
\bibitem [{\citenamefont {Nehl}\ \emph {et~al.}(2004)\citenamefont {Nehl},
  \citenamefont {Grady}, \citenamefont {Goodrich}, \citenamefont {Tam},
  \citenamefont {Halas},\ and\ \citenamefont {Hafner}}]{Nehl2004}%
  \BibitemOpen
  \bibfield  {author} {\bibinfo {author} {\bibfnamefont {C.~L.}\ \bibnamefont
  {Nehl}}, \bibinfo {author} {\bibfnamefont {N.~K.}\ \bibnamefont {Grady}},
  \bibinfo {author} {\bibfnamefont {G.~P.}\ \bibnamefont {Goodrich}}, \bibinfo
  {author} {\bibfnamefont {F.}~\bibnamefont {Tam}}, \bibinfo {author}
  {\bibfnamefont {N.~J.}\ \bibnamefont {Halas}}, \ and\ \bibinfo {author}
  {\bibfnamefont {J.~H.}\ \bibnamefont {Hafner}},\ }\href@noop {} {\bibfield
  {journal} {\bibinfo  {journal} {Nano Letters}\ }\textbf {\bibinfo {volume}
  {4}},\ \bibinfo {pages} {2355} (\bibinfo {year} {2004})}\BibitemShut
  {NoStop}%
\bibitem [{\citenamefont {Kirakosyan}\ \emph {et~al.}(2016)\citenamefont
  {Kirakosyan}, \citenamefont {Stockman},\ and\ \citenamefont
  {Shahbazyan}}]{Kirakosyan2016}%
  \BibitemOpen
  \bibfield  {author} {\bibinfo {author} {\bibfnamefont {A.~S.}\ \bibnamefont
  {Kirakosyan}}, \bibinfo {author} {\bibfnamefont {M.~I.}\ \bibnamefont
  {Stockman}}, \ and\ \bibinfo {author} {\bibfnamefont {T.~V.}\ \bibnamefont
  {Shahbazyan}},\ }\href@noop {} {\bibfield  {journal} {\bibinfo  {journal}
  {Physical Review B}\ }\textbf {\bibinfo {volume} {94}},\ \bibinfo {pages}
  {155429} (\bibinfo {year} {2016})}\BibitemShut {NoStop}%
\bibitem [{\citenamefont {Pellegrini}\ \emph {et~al.}(2016)\citenamefont
  {Pellegrini}, \citenamefont {Celebrano}, \citenamefont {Finazzi},\ and\
  \citenamefont {Biagioni}}]{Pellegrini2016}%
  \BibitemOpen
  \bibfield  {author} {\bibinfo {author} {\bibfnamefont {G.}~\bibnamefont
  {Pellegrini}}, \bibinfo {author} {\bibfnamefont {M.}~\bibnamefont
  {Celebrano}}, \bibinfo {author} {\bibfnamefont {M.}~\bibnamefont {Finazzi}},
  \ and\ \bibinfo {author} {\bibfnamefont {P.}~\bibnamefont {Biagioni}},\
  }\href@noop {} {\bibfield  {journal} {\bibinfo  {journal} {The Journal of
  Physical Chemistry C}\ }\textbf {\bibinfo {volume} {120}},\ \bibinfo {pages}
  {26021} (\bibinfo {year} {2016})}\BibitemShut {NoStop}%
\bibitem [{\citenamefont {Luther}\ \emph {et~al.}(2011)\citenamefont {Luther},
  \citenamefont {Jain}, \citenamefont {Ewers},\ and\ \citenamefont
  {Alivisatos}}]{Luther2011}%
  \BibitemOpen
  \bibfield  {author} {\bibinfo {author} {\bibfnamefont {J.~M.}\ \bibnamefont
  {Luther}}, \bibinfo {author} {\bibfnamefont {P.~K.}\ \bibnamefont {Jain}},
  \bibinfo {author} {\bibfnamefont {T.}~\bibnamefont {Ewers}}, \ and\ \bibinfo
  {author} {\bibfnamefont {A.~P.}\ \bibnamefont {Alivisatos}},\ }\href@noop {}
  {\bibfield  {journal} {\bibinfo  {journal} {Nature materials}\ }\textbf
  {\bibinfo {volume} {10}},\ \bibinfo {pages} {361} (\bibinfo {year}
  {2011})}\BibitemShut {NoStop}%
\bibitem [{\citenamefont {Koppens}\ \emph {et~al.}(2011)\citenamefont
  {Koppens}, \citenamefont {Chang},\ and\ \citenamefont {Garc{\'\i}a~de
  Abajo}}]{Koppens2011}%
  \BibitemOpen
  \bibfield  {author} {\bibinfo {author} {\bibfnamefont {F.~H.}\ \bibnamefont
  {Koppens}}, \bibinfo {author} {\bibfnamefont {D.~E.}\ \bibnamefont {Chang}},
  \ and\ \bibinfo {author} {\bibfnamefont {F.~J.}\ \bibnamefont {Garc{\'\i}a~de
  Abajo}},\ }\href@noop {} {\bibfield  {journal} {\bibinfo  {journal} {Nano
  letters}\ }\textbf {\bibinfo {volume} {11}},\ \bibinfo {pages} {3370}
  (\bibinfo {year} {2011})}\BibitemShut {NoStop}%
\bibitem [{\citenamefont {Mackowski}\ and\ \citenamefont
  {Mishchenko}(2011)}]{Mackowski2011}%
  \BibitemOpen
  \bibfield  {author} {\bibinfo {author} {\bibfnamefont {D.}~\bibnamefont
  {Mackowski}}\ and\ \bibinfo {author} {\bibfnamefont {M.}~\bibnamefont
  {Mishchenko}},\ }\href@noop {} {\bibfield  {journal} {\bibinfo  {journal}
  {Journal of Quantitative Spectroscopy and Radiative Transfer}\ }\textbf
  {\bibinfo {volume} {112}},\ \bibinfo {pages} {2182} (\bibinfo {year}
  {2011})}\BibitemShut {NoStop}%
\bibitem [{\citenamefont {Yang}\ \emph {et~al.}(2017)\citenamefont {Yang},
  \citenamefont {Miller}, \citenamefont {Christensen}, \citenamefont
  {Joannopoulos},\ and\ \citenamefont {Soljacic}}]{Yang2017}%
  \BibitemOpen
  \bibfield  {author} {\bibinfo {author} {\bibfnamefont {Y.}~\bibnamefont
  {Yang}}, \bibinfo {author} {\bibfnamefont {O.~D.}\ \bibnamefont {Miller}},
  \bibinfo {author} {\bibfnamefont {T.}~\bibnamefont {Christensen}}, \bibinfo
  {author} {\bibfnamefont {J.~D.}\ \bibnamefont {Joannopoulos}}, \ and\
  \bibinfo {author} {\bibfnamefont {M.}~\bibnamefont {Soljacic}},\ }\href@noop
  {} {\bibfield  {journal} {\bibinfo  {journal} {Nano Letters}\ }\textbf
  {\bibinfo {volume} {17}},\ \bibinfo {pages} {3238} (\bibinfo {year}
  {2017})}\BibitemShut {NoStop}%
\bibitem [{\citenamefont {Li}\ \emph {et~al.}(2017)\citenamefont {Li},
  \citenamefont {Fitzgerald}, \citenamefont {Xiao}, \citenamefont {Caldwell},
  \citenamefont {Zhang}, \citenamefont {Maier}, \citenamefont {Li},\ and\
  \citenamefont {Giannini}}]{Li2017}%
  \BibitemOpen
  \bibfield  {author} {\bibinfo {author} {\bibfnamefont {K.}~\bibnamefont
  {Li}}, \bibinfo {author} {\bibfnamefont {J.~M.}\ \bibnamefont {Fitzgerald}},
  \bibinfo {author} {\bibfnamefont {X.}~\bibnamefont {Xiao}}, \bibinfo {author}
  {\bibfnamefont {J.~D.}\ \bibnamefont {Caldwell}}, \bibinfo {author}
  {\bibfnamefont {C.}~\bibnamefont {Zhang}}, \bibinfo {author} {\bibfnamefont
  {S.~A.}\ \bibnamefont {Maier}}, \bibinfo {author} {\bibfnamefont
  {X.}~\bibnamefont {Li}}, \ and\ \bibinfo {author} {\bibfnamefont
  {V.}~\bibnamefont {Giannini}},\ }\href@noop {} {\bibfield  {journal}
  {\bibinfo  {journal} {ACS Omega}\ }\textbf {\bibinfo {volume} {2}},\ \bibinfo
  {pages} {3640} (\bibinfo {year} {2017})}\BibitemShut {NoStop}%
\bibitem [{\citenamefont {Johnson}\ and\ \citenamefont
  {Christy}(1972)}]{Johnson1972}%
  \BibitemOpen
  \bibfield  {author} {\bibinfo {author} {\bibfnamefont {P.~B.}\ \bibnamefont
  {Johnson}}\ and\ \bibinfo {author} {\bibfnamefont {R.-W.}\ \bibnamefont
  {Christy}},\ }\href@noop {} {\bibfield  {journal} {\bibinfo  {journal}
  {Physical review B}\ }\textbf {\bibinfo {volume} {6}},\ \bibinfo {pages}
  {4370} (\bibinfo {year} {1972})}\BibitemShut {NoStop}%
\bibitem [{\citenamefont {Blaber}\ \emph {et~al.}(2009)\citenamefont {Blaber},
  \citenamefont {Arnold},\ and\ \citenamefont {Ford}}]{Blaber2009}%
  \BibitemOpen
  \bibfield  {author} {\bibinfo {author} {\bibfnamefont {M.~G.}\ \bibnamefont
  {Blaber}}, \bibinfo {author} {\bibfnamefont {M.~D.}\ \bibnamefont {Arnold}},
  \ and\ \bibinfo {author} {\bibfnamefont {M.~J.}\ \bibnamefont {Ford}},\
  }\href@noop {} {\bibfield  {journal} {\bibinfo  {journal} {The Journal of
  Physical Chemistry C}\ }\textbf {\bibinfo {volume} {113}},\ \bibinfo {pages}
  {3041} (\bibinfo {year} {2009})}\BibitemShut {NoStop}%
\bibitem [{\citenamefont {Francescato}(2014)}]{Francescato2014}%
  \BibitemOpen
  \bibfield  {author} {\bibinfo {author} {\bibfnamefont {Y.}~\bibnamefont
  {Francescato}},\ }\emph {\bibinfo {title} {New frequencies and geometries for
  plasmonics and metamaterials}},\ \href@noop {} {Ph.D. thesis},\ \bibinfo
  {school} {Imperial College London} (\bibinfo {year} {2014})\BibitemShut
  {NoStop}%
\bibitem [{\citenamefont {Andrade}\ \emph {et~al.}(2015)\citenamefont
  {Andrade}, \citenamefont {Strubbe}, \citenamefont {De~Giovannini},
  \citenamefont {Larsen}, \citenamefont {Oliveira}, \citenamefont
  {Alberdi-Rodriguez}, \citenamefont {Varas}, \citenamefont {Theophilou},
  \citenamefont {Helbig}, \citenamefont {Verstraete} \emph
  {et~al.}}]{Andrade2015}%
  \BibitemOpen
  \bibfield  {author} {\bibinfo {author} {\bibfnamefont {X.}~\bibnamefont
  {Andrade}}, \bibinfo {author} {\bibfnamefont {D.}~\bibnamefont {Strubbe}},
  \bibinfo {author} {\bibfnamefont {U.}~\bibnamefont {De~Giovannini}}, \bibinfo
  {author} {\bibfnamefont {A.~H.}\ \bibnamefont {Larsen}}, \bibinfo {author}
  {\bibfnamefont {M.~J.}\ \bibnamefont {Oliveira}}, \bibinfo {author}
  {\bibfnamefont {J.}~\bibnamefont {Alberdi-Rodriguez}}, \bibinfo {author}
  {\bibfnamefont {A.}~\bibnamefont {Varas}}, \bibinfo {author} {\bibfnamefont
  {I.}~\bibnamefont {Theophilou}}, \bibinfo {author} {\bibfnamefont
  {N.}~\bibnamefont {Helbig}}, \bibinfo {author} {\bibfnamefont {M.~J.}\
  \bibnamefont {Verstraete}},  \emph {et~al.},\ }\href@noop {} {\bibfield
  {journal} {\bibinfo  {journal} {Physical Chemistry Chemical Physics}\
  }\textbf {\bibinfo {volume} {17}},\ \bibinfo {pages} {31371} (\bibinfo {year}
  {2015})}\BibitemShut {NoStop}%
\end{thebibliography}%

\end{document}